\documentclass[oldversion]{aa}  

\usepackage{graphicx}
\usepackage{txfonts}
\usepackage{xcolor}
\usepackage{ragged2e}
\newcommand{\hi}{{\sc H\,i}}
\newcommand{\HIbf}{\mbox{H\hspace{0.155 em}{\footnotesize \bf I}}}

\newcommand{\mhi}{$M_\mathrm{HI}$}

\newcommand{\mstar}{$M_\star$}
\newcommand{\Nhi}{$N$(\hi)}
\newcommand{\mJybeam}{mJy beam$^{-1}$}
\newcommand{\msun}{{M$_\odot$}}

\newcommand{\kms}{$\,$km$\,$s$^{-1}$}
\newcommand{\ltsima} {$\; \buildrel < \over \sim \;$}
\newcommand{\gtsima} {$\; \buildrel > \over \sim \;$}
\newcommand{\lta} {\lower.5ex\hbox{\ltsima}}
\newcommand{\gta} {\lower.5ex\hbox{\gtsima}}

\newcommand{\Ha}{H$\alpha$}

\begin{document}

   \title{The MeerKAT Fornax Survey. III. Ram-pressure stripping of the tidally interacting galaxy NGC~1427A in the Fornax cluster}
   \titlerunning{The MeerKAT Fornax Survey - III}

   \author{
   P.~Serra \inst{1}\fnmsep\thanks{paolo.serra@inaf.it},
   T.~A.~\rm Oosterloo \inst{2,3},
   P.~Kamphuis \inst{4},
   G.~I.~G.~J\'{o}zsa \inst{5,6},
   W.~J.~G.~de~Blok \inst{2,3,7},
   G.~L.~Bryan \inst{8},
   J.~H.~van~Gorkom \inst{8},
   E.~Iodice \inst{9},
   D.~Kleiner \inst{2},
   A.~Loni \inst{9},
   S.~I.~Loubser \inst{10},
   F.~M.~Maccagni \inst{1},
   D.~Moln\'{a}r \inst{1},
   R.~Peletier \inst{3},
   D.~J.~Pisano \inst{7},
   M.~Ramatsoku \inst{6,1},
   M.~W.~L.~Smith \inst{11},
   M.~A.~W.~Verheijen \inst{3},
   \and
   N.~Zabel \inst{7}
          }
          
   \authorrunning{Serra et al.}

   \institute{
   INAF - Osservatorio Astronomico di Cagliari, Via della Scienza 5, I-09047 Selargius (CA), Italy
   \and
   Netherlands Institute for Radio Astronomy (ASTRON), Oude Hoogeveensedijk 4, 7991 PD Dwingeloo, the Netherlands
   \and
   Kapteyn Astronomical Institute, University of Groningen, PO Box 800, NL-9700 AV Groningen, the Netherlands
   \and
   Ruhr University Bochum, Faculty of Physics and Astronomy, Astronomical Institute (AIRUB), 44780 Bochum, Germany
   \and
   Max-Planck-Institut f\"{u}r Radioastronomie, Auf dem H\"{u}gel 69, D-53121 Bonn, Germany
   \and
   Department of Physics and Electronics, Rhodes University, PO Box 94, Makhanda 6140, South Africa
   \and
   Department of Astronomy, University of Cape Town, Private Bag X3, Rondebosch 7701, South Africa
   \and
   Department of Astronomy, Columbia University, New York, NY 10027, USA
   \and
   INAF - Astronomical Observatory of Capodimonte, Salita Moiariello 16, 80131, Naples, Italy
   \and
   Centre for Space Research, North-West University, Potchefstroom 2520, South Africa
   \and
   School of Physics and Astronomy, Cardiff University, Queens Buildings, The Parade, Cardiff CF24 3AA, UK
   \\
             }

   \date{Received ...; accepted ...}

  \abstract
  {We present MeerKAT Fornax Survey \hi\ observations of NGC~1427A, a blue irregular galaxy with a stellar mass of $\sim2\times10^9$ \msun\ located near the centre of the Fornax galaxy cluster. Thanks to the excellent resolution (1 to 6 kpc spatially, 1.4 \kms\ in velocity) and \hi\ column density sensitivity ($\sim4\times10^{19}$ to $\sim10^{18}$ cm$^{-2}$ depending on resolution), our data deliver new insights on the long-debated interaction of this galaxy with the cluster environment. We confirm the presence of a broad, one-sided, starless \hi\ tail stretching from the outer regions of the stellar body and pointing away from the cluster centre. We find the tail to have 50\% more \hi\ ($4\times10^8$ \msun) and to be 3 times longer (70 kpc) than in previous observations. In fact, we detect scattered \hi\ clouds out to 300 kpc from the galaxy in the direction of the tail --- possibly the most ancient remnant of the passage of NGC~1427A through the intracluster medium of Fornax. Both the velocity gradient along the \hi\ tail and the peculiar kinematics of \hi\ in the outer region of the stellar body are consistent with the effect of ram pressure given the line-of-sight motion of the galaxy within the cluster. However, several properties cannot be explained solely by ram pressure and suggest an ongoing tidal interaction. This includes: the close match between dense \hi\ and stars within the disturbed stellar body; the abundant kinematically-anomalous \hi; and the inversion of the \hi\ velocity gradient near the base of the \hi\ tail. We rule out an interaction with the cluster tidal field, and conclude that NGC~1427A is the result of a high-speed galaxy encounter or of a merger started at least 300 Myr ago, where ram pressure shapes the distribution and kinematics of the \hi\ in the perturbed outer stellar body and in the tidal tails.}
     

   \maketitle

\section{Introduction}
\label{sec:intro}

The systematic change of basic galaxy properties such as morphology, cold gas content and star formation activity as a function of position in the cosmic web has been known for decades and continues to be refined through new observations \citep[e.g.,][]{hubble1931,spitzer1951,oemler1974,dressler1980,larson1980,giovanelli1983,postman1984,vader1991,whitmore1993,fasano2000,koopman2004,postman2005,vanderwel2007,cappellari2011b,bait2017,brown2017,shimakawa2021}. Several mechanisms are thought to contribute to the emergence and evolution of this relation throughout cosmic time. In low-density environments (e.g., large-scale filaments or loose groups) relatively weak and/or rare tidal and hydrodynamical interactions may hamper the accretion and cooling of gas from the intergalactic and circumgalactic medium. In such cases the interstellar medium (ISM) is consumed through star formation without being replenished, leading to the slow decline of the star formation rate \citep[e.g.,][]{larson1980}. In higher-density environments (e.g., massive groups and galaxy clusters) stronger and/or more frequent interactions are able to directly strip galaxies of their ISM and thus quench star formation more rapidly \citep[e.g.,][]{gallagher1972a,gunn1972,cowie1977,nulsen1982,moore1996,bekki1999}. The abundant observational evidence of stripping in dense environments and its impact on galaxy evolution have been reviewed most recently by \cite{cortese2021} and \cite{boselli2022}.

Despite the steady progress, the exact balance between the processes leading to slow and fast quenching, and the balance between tidal and hydrodynamical interactions as a function of environment and galaxy properties remain difficult to establish. An important open question is whether hydrodynamical stripping of the ISM --- a key process in massive galaxy clusters --- can actually happen in smaller clusters and groups, below a virial mass of $\sim10^{14}$ \msun, and how this depends on galaxy properties. In this context, nearby, low-mass clusters represent excellent targets for detailed observational studies. Among them, the nearest is Fornax, which has a virial mass of $\sim5\times10^{13}$ \msun\ \citep{drinkwater2001b} and lies at a distance of $\sim20$ Mpc \citep{blakeslee2001,blakeslee2009,jensen2001,tonry2001}.

\begin{figure}
\includegraphics[width=9cm]{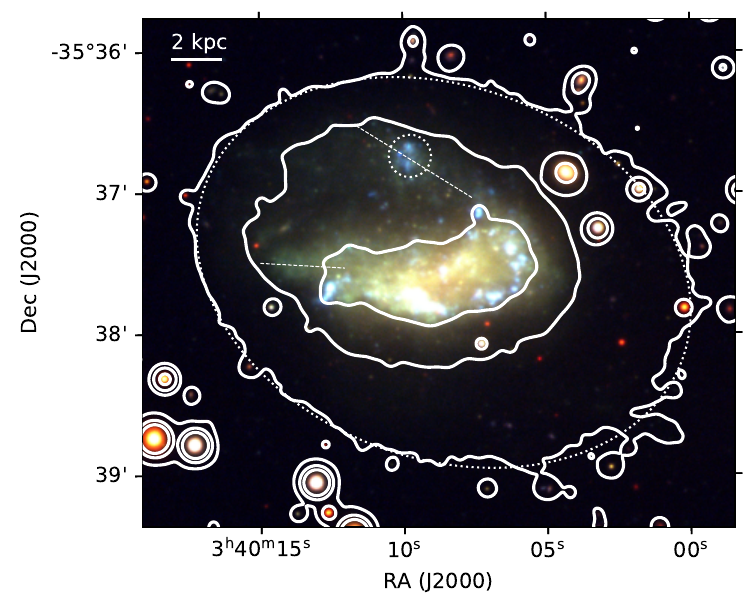}
\caption{RGB image of NGC~1427A made using the Fornax Deep Survey images \citep{iodice2016,venhola2018} in band $i$ (R channel), $g$ (G channel) and $u$ (B channel). The solid white contours are three isophotes from the $g$-band image convolved with a $5''$-FWHM Gaussian kernel. The isophotes are spaced by a factor of 5 in surface brightness ($\sim1.75$ mag arcsec$^{-2}$) and highlight regions of the stellar body described in Sect. \ref{sec:intro}: \emph{i)} the inner-most isophote shows the bright stellar body of NGC~1427A with a semi-major axis of $50''$ ($\sim5$ kpc) and PA $\sim 100^\circ$; \emph{ii)} the middle isophote is asymmetric relative to the centre of the bright stellar body and includes the northern clump (dotted white circle), the northern plume (northern dashed line) and a faint elongated feature stretching from the eastern edge of the bright stellar body (southern dashed line); \emph{iii)} the faintest isophote has an elliptical shape centred on the bright part of the stellar body, with a semi-major axis of $1.8'$ ($\sim10$ kpc) and PA $\sim70^\circ$ (dotted white ellipse).}
\label{fig:opt}
\end{figure}

The MeerKAT Fornax Survey\footnote{https://sites.google.com/inaf.it/meerkatfornaxsurvey} \citep{serra2023} is producing unprecedentedly sensitive neutral hydrogen (\hi) cubes and images of the Fornax cluster at resolutions from $\sim10''$ ($\sim1$ kpc at its 20 Mpc distance) to $\sim100''$ ($\sim10$ kpc). \hi\ is an excellent tracer of the interaction between galaxies and their environment. Indeed, using the $40''$-resolution \hi\ data of this survey, \cite{serra2023} presented the first, unambiguous evidence of ram pressure shaping the distribution of \hi\ in this cluster: a sample of six galaxies with long, one-sided, radially-oriented, starless \hi\ tails, whose velocity gradient is consistent with the ram-pressure caused by the line-of-sight motion of the host galaxies within the cluster. \cite{serra2023} also suggested that, given the signs of recent tidal interactions in all galaxies with an \hi\ tail, it is possible that the \hi\ was pulled away from the stellar body by tidal forces and, once there, was further displaced by the relatively weak ram pressure of Fornax. 

Although they demonstrated that ram pressure plays an important role in shaping the distribution of the \hi\ in the Fornax cluster, \cite{serra2023} did not discuss the balance between tidal and hydrodynamical effects in detail, nor did they establish how deep into the stellar body ram pressure can affect the ISM. Further progress requires a more detailed analysis of the \hi\ distribution and kinematics of those Fornax galaxies. In this paper we present such analysis for one of them.

   \begin{table}
   {\centering
      \caption[]{Main properties of NGC~1427A}
         \label{table:n1427a}
         \begin{tabular}{lll}
            \noalign{\smallskip}
            \hline
            \hline
            \noalign{\smallskip}

            IDs &  \multicolumn{2}{l}{NGC~1427A, ESO~358-G049, FCC~235} \\
            RA (J2000)$^{(1)}$ & 3h 40m 8.7s & this work \\
            Dec (J2000)$^{(1)}$ & $-35$d 37m 34s & this work \\
            $v_\mathrm{sys}\ ^{(1)}$ & 2036 \kms\ & this work \\
            distance$^{(2)}$ & 20 Mpc & \cite{tonry2001} \\
            \mstar & $(2.3\pm0.5)\times10^9$ \msun\ & \cite{loni2021} \\
            \mhi$^{(3)}$ & $(2.3\pm0.2)\times10^9$ \msun\ & this work \\

            \noalign{\smallskip}
            \hline
            \noalign{\smallskip}
         \end{tabular}

         \justifying
         \noindent \small \emph{Notes.} (1) The central coordinates and systemic velocity are those of the dynamical centre (Sect. \ref{sec:dense}). Throughout this paper we list barycentric velocities in the optical convention $v = c\times(\nu_0/\nu-1)$, where $c$ is the speed of light, $\nu_0$ the rest frequency, and $\nu$ the observed frequency in the barycentric frame of reference. (2) This is the distance of the Fornax cluster as a whole. (3) The value reported here for the \hi\ mass is measured from the image at $66''$ resolution, while at higher resolution it is slightly lower. The uncertainty is calculated by adding in quadrature the statistical uncertainty within the \hi\ detection mask (which depends on the noise level in the cube and the number of independent voxels in the mask) and a 10\% flux scale uncertainty \citep{serra2023}. The latter term dominates the error budget.
         
         }
   \end{table}

NGC~1427A is a blue irregular galaxy similar to the Large Magellanic Cloud (see Fig. \ref{fig:opt} and Table \ref{table:n1427a}). Its peculiar morphology was first noted by \cite{laustsen1987} and since then studied at various wavelengths by \cite{cellone1997}, \cite{hilker1997}, \cite{chaname2000}, \cite{georgiev2006}, \cite{sivanandam2014}, \cite{mora2015}, \cite{leewaddell2018} and \cite{mastropietro2021}. With reference to Fig. \ref{fig:opt}, the bright part of the stellar body has an elongated shape with position angle PA $\sim100^\circ$ out to a semi-major axis of $50''$ ($\sim5$ kpc; see the brightest isophote in the figure). The south-west (SW) edge of the bright stellar body is dotted with star-forming regions, which are also detected in ionised gas through the \Ha\ line \citep{chaname2000,georgiev2006}. Further out, the distribution of stellar light becomes less regular: an elongated feature (southern dashed line in the figure) stretches from the eastern end of the bright stellar body towards east; another one (northern dashed line, hereafter called ``northern plume'') goes from the western end towards north-east (NE) and includes a bright, blue clump (dotted circle, hereafter called ``northern clump''). This clump is also detected in \Ha\ \citep{chaname2000,georgiev2006}, suggesting ongoing star formation, and has been discussed as a possible merging companion \citep{cellone1997}. Together, these features result in a marked asymmetry of the middle isophote in Fig. \ref{fig:opt}, with an overall shift towards NE. Finally, the faintest isophote in the figure has a fairly regular elliptical shape with a semi-major axis of $1.8'$ ($\sim10$ kpc) and PA $\sim 70^\circ$ (dotted ellipse in the figure). Its centre is shifted back towards SW compared to that of the middle isophote and is remarkably coincident with the overall centre of the galaxy (Table \ref{table:n1427a}). This shifting of the centre of the isophotes as a function of radius is a further confirmation that the stellar body is unsettled.

NGC~1427A has attracted much attention because of the complex distribution of star forming regions, young star clusters and warm molecular gas along the SW edge of the bright stellar body. Based on these observations, \cite{chaname2000}, \cite{sivanandam2014} and \cite{mora2015} proposed that the ISM in this region is compressed by the ram pressure exerted by the intracluster medium (ICM) of Fornax as NGC~1427A moves towards SW within it.

At face value, NGC~1427A would indeed appear as an excellent candidate for ram-pressure effects in Fornax. First, it is located just $\sim 130$ kpc south-east (SE) of the cluster central galaxy NGC~1399 in projection ($\sim0.2\times$ the virial radius; \citealt{drinkwater2001b})\footnote{The distance along the line of sight is highly uncertain: NGC~1427A is formally in the foreground of NGC~1399 by 3.2 Mpc $\pm2.5$ Mpc (statistic) $\pm1.6$ Mpc (systematic) based on the globular cluster luminosity function of the two galaxies \citep{georgiev2006}. This is consistent with the galaxy being within the cluster. The fact that NGC~1427A is strongly red-shifted relative to the cluster (see text) makes it unlikely that it is significantly in its foreground, as this would imply an extraordinarily high infall speed upon entering Fornax.}. Therefore, NGC~1427A might be moving through a region where the ICM density is relatively high: $n_\mathrm{ICM}\sim10^{-4}$ cm$^{-3}$ according to the density profile from \cite{paolillo2002}. Furthermore, NGC~1427A moves at high speed within the cluster: $v = 584$ \kms\ relative to NGC~1399 along the line of sight ($\sim1.8\times$ the velocity dispersion of Fornax; \citealt{maddox2019}). Assuming that the 3D velocity is $\sqrt{3}\times v$ the resulting ram pressure is $m_\mathrm{p} \ n_\mathrm{ICM}\  3 v^2\sim1.8\times10^{-12}$ erg cm$^{-3}$, where $m_\mathrm{p}$ is the proton mass \citep{gunn1972}. For comparison, the \emph{average} gravitational restoring force per unit area within a 5 kpc radius (which encloses about \mstar/2 and where the ISM surface mass density is $\Sigma_\mathrm{ISM}\sim10$ \msun\ pc$^{-2}$; see Table \ref{table:n1427a} and Sect. \ref{sec:results}) is $2\pi \ G \ \Sigma_\star \ \Sigma_\mathrm{ISM}\sim3.4\times10^{-12}$ erg cm$^{-3}$, where $G$ is the gravitational constant and $\Sigma_\star$ the stellar surface mass density. Therefore, ram pressure might be sufficient to affect the ISM \emph{at least} at the outskirts of NGC~1427A.

Despite this, recent results demonstrate that ram pressure cannot be the dominant cause of the SW star-forming front. If it were, we would expect a tail of stripped ISM directed towards NE on the opposite side of the galaxy. However, the first \hi\ image with the resolution and sensitivity required to detect such a tail revealed  a 25-kpc-long \hi\ tail starting south of the stellar body and pointing SE, \emph{perpendicular} to the expected direction \citep{leewaddell2018}. This discovery demonstrated that \emph{if} ram pressure is acting on NGC~1427A then it must be pushing the gas towards SE and cannot be at the origin of the SW star-forming front. More recently, \cite{serra2023} showed that the SE tail is actually much longer, that it is starless and radially aligned within the cluster across its entire length, and that the \hi\ velocity gradient along the tail is consistent with the blue-shifting effect of ram-pressure along the line of sight (NGC~1427A is red-shifted relatively to the cluster centre).

\begin{figure*}
\centering
\includegraphics[width=18cm]{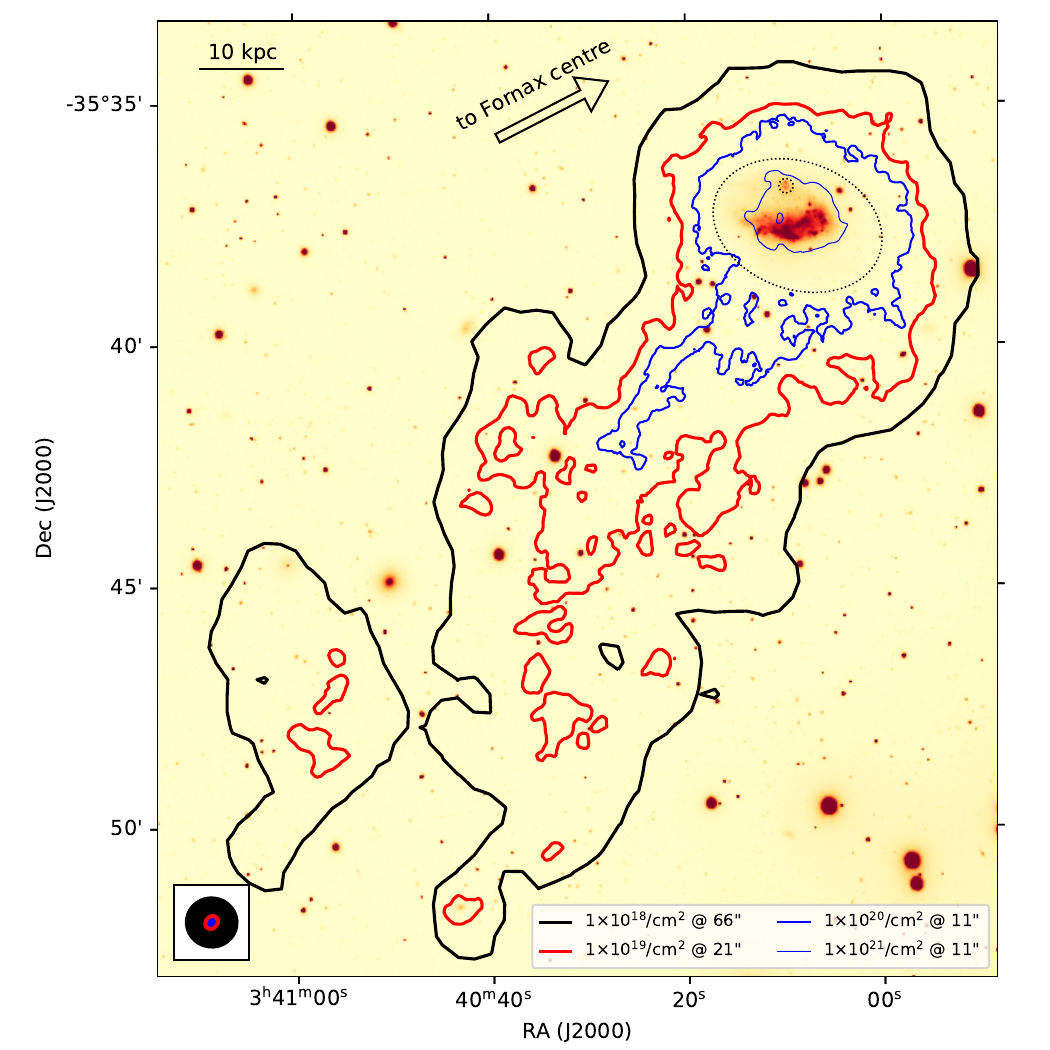}
\caption{MeerKAT \hi\ contours overlaid on a $g$-band optical image from the Fornax Deep Survey. The contours are drawn at different levels and resolutions as indicated in the legend on the bottom-right. The corresponding beam sizes are shown in the bottom-left corner. The dotted circle and ellipse represent the northern clump and the outer isophote of the stellar body, respectively (Fig. \ref{fig:opt}).}
\label{fig:hicontours}
\end{figure*}

The results presented by \cite{serra2023} demonstrate that ram pressure is indeed shaping the SE \hi\ tail. The peculiar optical appearance of NGC~1427A must then be caused by some effects other than ram pressure. The obvious candidate is a tidal interaction. The idea of a galaxy merger was already discussed by \cite{cellone1997} and then again by \cite{leewaddell2018} on the basis of their \hi\ observations. In particular, the aforementioned northern clump (white dotted circle in Fig. \ref{fig:opt}) could be the remnant of an intruder which disturbed the main progenitor --- though the clump's kinematics does not appear to be anomalous \citep{chaname2000}. Alternatively, \cite{mastropietro2021} proposed that NGC~1427A is distorted by the tidal field of the Fornax cluster as the galaxy approaches pericentre along a radial orbit. In both cases, ram pressure might be acting simultaneously with the tidal forces and form the observed, one-sided southern \hi\ tail, as argued by \cite{serra2023}.

The jury is still out on the exact sequence of events and the physical mechanisms that are shaping the properties of NGC~1427A inside Fornax. Having established through moderate-resolution \hi\ imaging that ram pressure \it i) \rm is not causing the SW star-forming front \citep{leewaddell2018} and \it ii) \rm is shaping the southern \hi\ tail \citep{serra2023}, what do higher-resolution data tell us about the recent evolution of this galaxy? How deep into the stellar body is the ISM being displaced by ram pressure? Are tidal forces also at play and, if so, are they due to a galaxy merger or to an interaction with the cluster potential? And, finally, what is the interplay between --- and the combined effect of --- tidal forces and ram pressure?

In this paper we attempt to answer these questions through multi-resolution \hi\ observations of NGC~1427A obtained as part of the MeerKAT Fornax Survey. We describe the data in Sect. \ref{sec:data}, show the results in Sect. \ref{sec:results}, discuss our interpretation in Sect. \ref{sec:interpr} and summarise our conclusions in Sect. \ref{sec:summary}.

\section{MeerKAT data}
\label{sec:data}

\begin{figure}
\centering
\includegraphics[width=9cm]{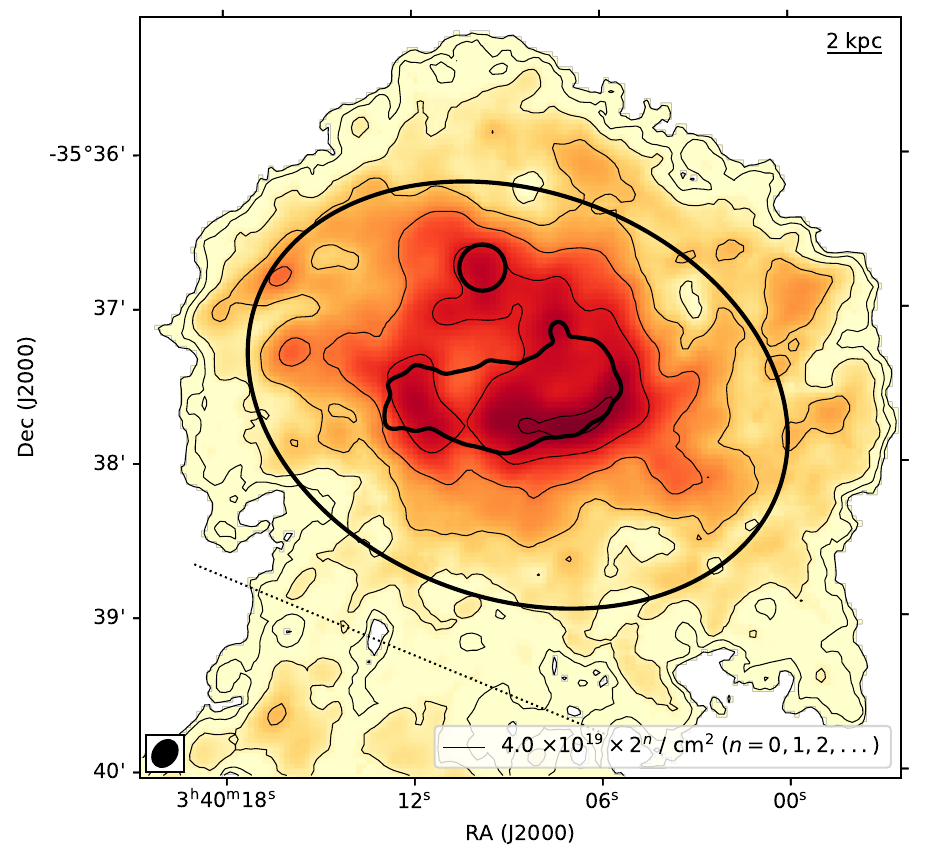}
\caption{Total \hi\ image at an angular resolution of $11''$ with, overlaid in black, the main optical components of NGC~1427A: the bright stellar body, the northern clump, and the outer optical isophote (Fig. \ref{fig:opt}). The start of the \hi\ tail is visible south of the dotted line.}
\label{fig:tot11}
\end{figure}

We make use of the \hi\ data products described in \cite{serra2023}. In particular, we analyse the data cubes and moment images with an angular resolution of $11''$, $21''$, and $66''$, approximately corresponding to 1, 2 and 6 kpc, respectively, at the 20 Mpc distance of Fornax (for this galaxy, data at even lower resolution do not add significant information). In all cases the velocity resolution is 1.4 \kms. Locally, the noise level of the cubes can be slightly different from the average value reported by \cite{serra2023} for the wider Fornax field (see their Table 2). In the case of NGC~1427A it is 0.24, 0.26, and 0.29 \mJybeam\ at a resolution of $11''$, $21''$ and $66''$, respectively. The corresponding formal $3\sigma$ \hi\ column density sensitivity calculated over a 25 \kms\ line width is $4.0\times10^{19}$, $1.1\times10^{19}$ and $1.3\times10^{18}$ cm$^{-2}$, respectively.

\section{Results: \HIbf\ in NGC~1427A}
\label{sec:results}

Figure \ref{fig:hicontours} shows our multi-resolution MeerKAT view of \hi\ in NGC~1427A. The figure displays the considerable extent of the \hi\ distribution relative to the stellar body and the wide range of \hi\ column densities detectable in our data (more than 3 orders of magnitude). At the lowest resolution ($66''$) we measure a total \hi\ flux of $(24\pm2)$ Jy \kms, in agreement with previous estimates (\citealt{bureau1996,koribalski2004,leewaddell2018,loni2021}; see Table \ref{table:n1427a} for the calculation of the uncertainty). Using standard \hi\ conversions \citep{meyer2017} this corresponds to \mhi\ $=(2.3\pm0.2)\times10^9$ \msun\ at the assumed distance of 20 Mpc.

Figure \ref{fig:hicontours} highlights the complexity of the \hi\ distribution within and around NGC~1427A. At \Nhi\ $\sim10^{21}$ cm$^{-2}$ (thin blue contour) the gas traces the high-surface-brightness part of the stellar body as well as the northern plume centred on the northern clump (black dotted circle; see Sect. \ref{sec:intro} and Fig. \ref{fig:opt}). At \Nhi\ $\sim10^{20}$ cm$^{-2}$ (thick blue contour) the gas distribution has a similar ellipticity and PA as the outer optical isophote of Fig. \ref{fig:opt} (here represented by a black dotted ellipse) while extending just beyond it. Furthermore, at this column density the southern \hi\ tail begins to be visible. At  \Nhi\ between $10^{18}$ and $10^{19}$ cm$^{-2}$ (black and red contours, respectively) the full length and width of the tail become apparent.

The strong connection between dense \hi\ and bright optical emission can be appreciated in Fig. \ref{fig:tot11}, where we zoom into the $11''$-resolution \hi\ image. The densest gas is found in regions rich in star forming complexes (Fig. \ref{fig:opt}), both at the southern edge of the bright stellar body and within the northern clump. Furthermore, dense gas is found along the northern plume connecting the northern clump itself with the western edge of the bright stellar body. Finally, the figure confirms that \hi\ at $\sim10^{20}$ cm$^{-2}$ is distributed similarly to the outer optical isophote, although there appears to be an excess of \hi\ with a column density above $\sim10^{20}$ cm$^{-2}$ just outside the outer optical isophote in the NW quadrant. Similar to Fig. \ref{fig:hicontours}, also in Fig. \ref{fig:tot11} the main difference between optical and \hi\ morphology is the presence of the southern \hi\ tail.

\begin{figure}
\centering
\includegraphics[width=9cm]{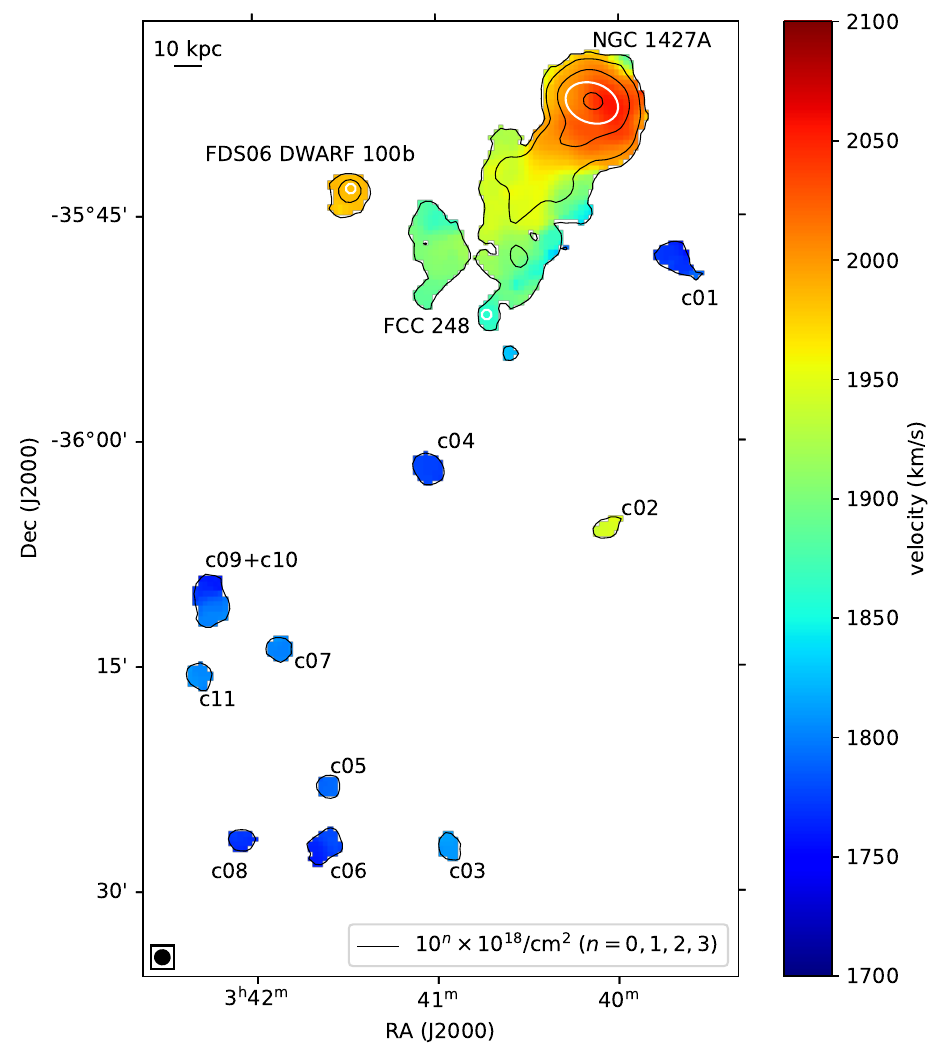}
\caption{Moment-1 \hi\ velocity field of the $66''$-resolution cube, showing all gas located within and near NGC~1427A both spatially and in velocity. The black contours represent the \hi\ image at the same resolution, with contour levels indicated in the bottom-right corner. The bottom-left black ellipse represents the $66''$ resolution. The large white ellipse represents the outer isophote of the stellar body (Fig. \ref{fig:opt}). Two more \hi-detected dwarf galaxies are present in this field: FCC~248 and FDS06~DWARF~100b. Their position is indicated by small white circles. The \hi\ clouds discussed in the text are identified by the same labels used in Fig. \ref{fig:clouds}.}
\label{fig:vfwide}
\end{figure}

\begin{figure}
\centering
\includegraphics[width=9cm]{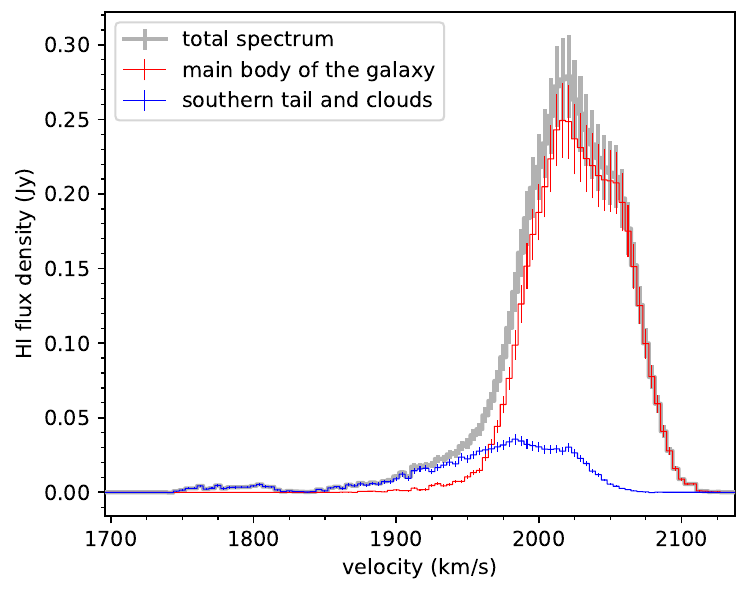}
\caption{\hi\ spectrum of NGC~1427A obtained integrating the $66''$-resolution cube within the detection mask and binning along the spectral axis by a factor of 3, resulting in a channel width of 4.1 \kms. The error bars are the sum in quadrature of the statistical error (derived from the noise level of the cube and the number of spatial pixels included in the integral of each channel) and a 10\% flux scale uncertainty \citep{serra2023}. The spectrum is in good agreement with previously published spectra \citep{koribalski2004,loni2021}. We refer to the text for a description of the decomposition of the total spectrum (grey line) in two components (red and blue line, respectively).}
\label{fig:1dspec}
\end{figure}

Zooming out, Fig. \ref{fig:vfwide} shows the spatial distribution and the line-of-sight velocity of the \hi\ in a relatively wide field around NGC~1427A. We detect \hi\ outside of this field, too, but we do not show it here because it has a widely different velocity and is associated with other galaxies in Fornax, with no signs of interaction with our target. The decrease of the \hi\ velocity towards SE along the southern tail was already described by \cite{serra2023}, who noted it to be consistent with the direction of the ram-pressure wind felt by the galaxy because of its line-of-sight motion through the ICM. The new result here is that, following the direction of the tail past its edge, we find a number of additional \hi\ clouds whose low velocity is broadly consistent with the velocity gradient along the tail. Figure \ref{fig:1dspec} shows the total \hi\ spectrum of the field (excluding the two dwarf galaxies labelled in Fig. \ref{fig:vfwide} and discussed below). We divide the spectrum in two components: the main \hi\ body, approximately defined as the region north of the dotted line shown in Fig. \ref{fig:tot11}; and the tail and clouds, south of the line. The clouds dominate the faint part of the spectrum at velocities between $\sim 1750$ and $\sim 1850$ \kms, while the division between the tail and the main \hi\ body is discussed in detail in Sect. \ref{sec:tails}.

\begin{figure}
\centering
\includegraphics[width=4.4cm]{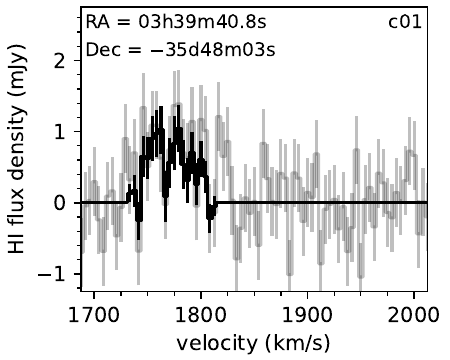}
\includegraphics[width=4.4cm]{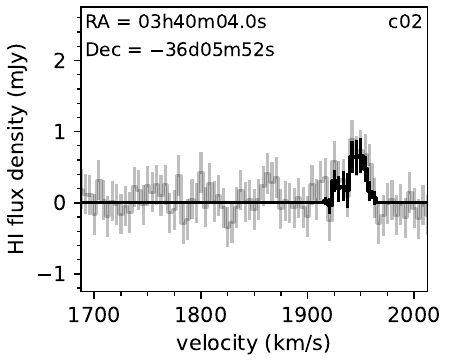}
\includegraphics[width=4.4cm]{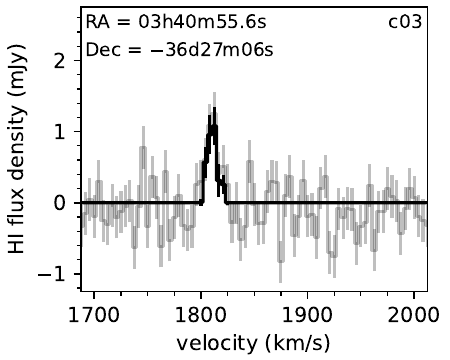}
\includegraphics[width=4.4cm]{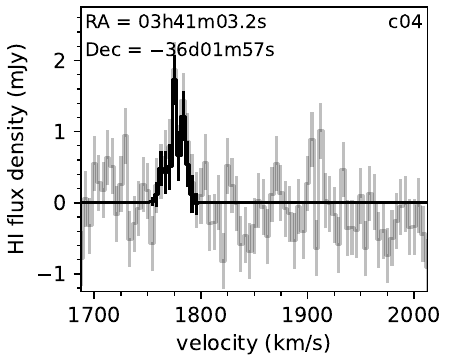}
\includegraphics[width=4.4cm]{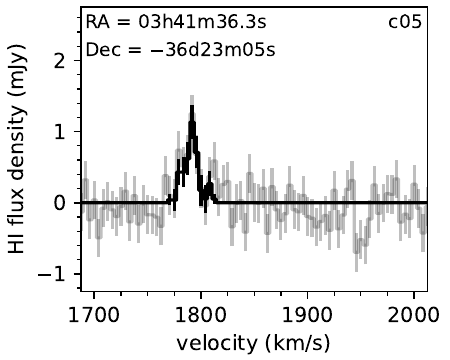}
\includegraphics[width=4.4cm]{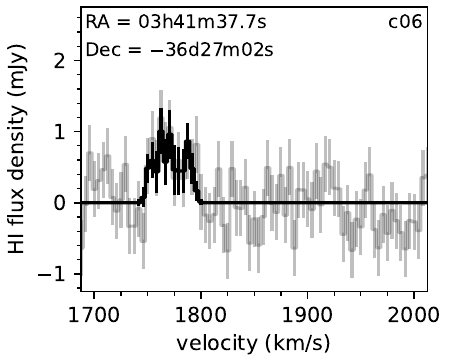}
\includegraphics[width=4.4cm]{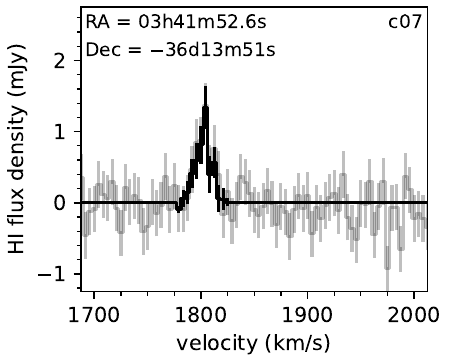}
\includegraphics[width=4.4cm]{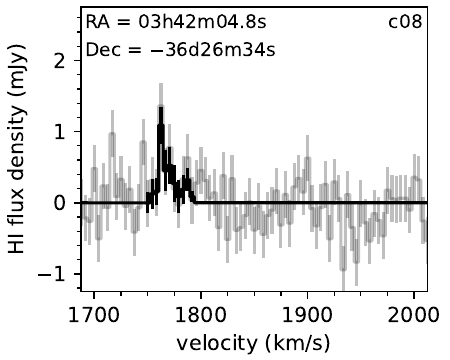}
\includegraphics[width=4.4cm]{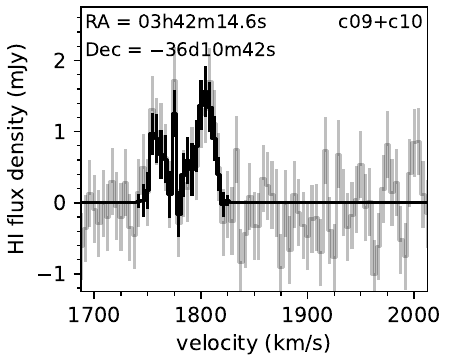}
\includegraphics[width=4.4cm]{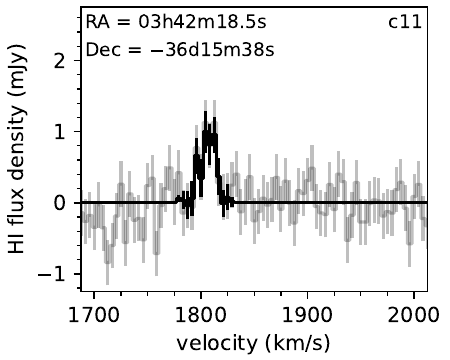}
\caption{Integrated \hi\ spectrum of the ten clouds shown in Fig. \ref{fig:vfwide} sorted in order of increasing RA (left to right, top to bottom) and using the same labels as in that figure. The spectra are obtained from the $66''$-resolution cube and are binned by a factor of 3 along the spectral axis, resulting in a channel width of 4.1 \kms. For each cloud, the black line shows the spectrum integrated within the 3D detection mask, while the grey line shows the spectrum integrated over a 2D aperture obtained collapsing the detection mask along the spectral axis. In both cases, we show statistical error bars based on the local noise level and on the number of spatial pixels included in the integral of each channel. We show the position of the \hi\ centre of mass of each cloud in the top-left corner. The bottom-left panel shows two clouds that are blended in 3D.}
\label{fig:clouds}
\end{figure}

Figure \ref{fig:clouds} shows the integrated spectrum of the individual clouds (including the two westernmost clouds visible in Fig. \ref{fig:vfwide} SW and south of NGC~1427A, respectively, which seem unrelated to the tail). The integrated  signal-to-noise ratio of the clouds ranges between 5 and 7, which makes their detection solid but their morphology poorly constrained. The clouds have a typical \hi\ mass between $\sim10^6$ and $\sim3\times10^6$ \msun. None of them has an optical counterpart down to a surface brightness of $\sim28$ mag arcsec$^{-2}$ (at $1''$ resolution) in the deep optical images of the Fornax Deep Survey \citep{iodice2016,venhola2018}. The clouds could thus be the diffuse, clumpy, extreme extension of the \hi\ tail of NGC~1427A --- the most ancient remnant of the passage of this galaxy through the ICM. Altogether, the \hi\ clouds in Fig. \ref{fig:vfwide} account for a mere 1\% of the total \hi\ mass of NGC~1427A.

Finally, two \hi-detected dwarf galaxies are visible in Fig. \ref{fig:vfwide}: FCC~248 (also visible at the bottom of Fig. \ref{fig:hicontours}) and FDS06~DWARF~100b. Both are discussed by \cite{kleiner2023}, who noted that the gas content of FCC~248 is uncertain given its projection against the southern end of the \hi\ tail and the lack of an optical redshift. We find no indications of a possible interaction between these dwarfs and NGC~1427A in our \hi\ data and in the deep optical images of the Fornax Deep Survey. Furthermore, both galaxies are more than 5 magnitudes fainter than NGC~1427A in optical bands \citep{su2022} and are therefore too small to be responsible for the peculiar \hi\ morphology and kinematics of our target. For these reasons, we do not discuss these galaxies further. In the rest of this section we describe the \hi\ distribution and kinematics of NGC~1427A in more detail.

\subsection{Southern \hi\ tail}
\label{sec:tails}

\begin{figure*}
\includegraphics[width=18cm]{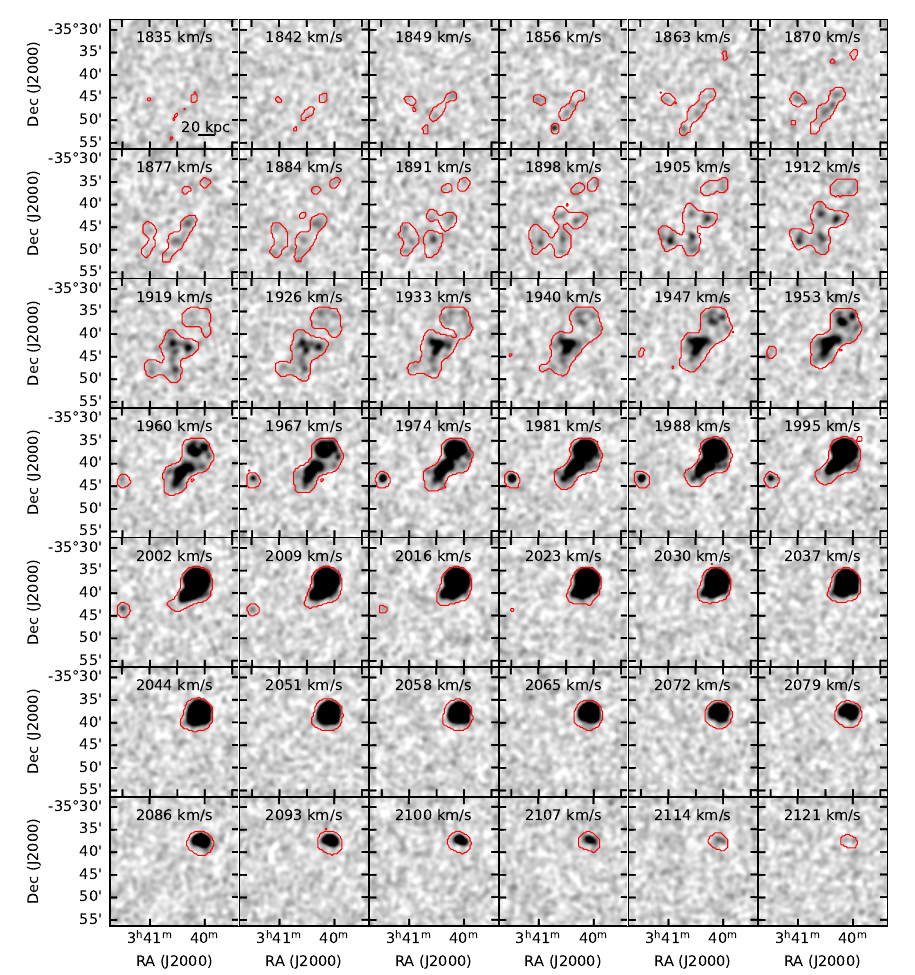}
\caption{Channel maps of the $66''$-resolution \hi\ cube of NGC~1427A. For visualisation purpose we binned the channels by a factor of 5 compared to the specs given in Sect. \ref{sec:data}, resulting in a channel width of 6.9 \kms. We use a linear grey scale from $-2\sigma$ to $+10\sigma$ in order to emphasise the low surface brightness emission, where $\sigma$ is the noise level. The red contour represents the detection mask used to make the \hi\ moment images.}
\label{fig:t60chans}
\end{figure*}

Figure \ref{fig:t60chans} shows the channel maps of the $66''$-resolution \hi\ cube binned by a factor of 5 in velocity for visualisation purpose, resulting in a channel width of 6.9 \kms. The angular and velocity resolution of these maps are similar to those of the previous best \hi\ observation of this galaxy obtained with the Australia Telescope Compact Array by \cite{leewaddell2018}. The comparison to their channel maps (their Figure 1) is therefore straightforward. The clearest new result at this resolution concerns the southern \hi\ tail first detected by those authors. In their data the tail stops at Dec $\sim -35$d 42m. In our new MeerKAT data, thanks to the $\sim25\times$ better \Nhi\ sensitivity, the tail appears much longer, reaching Dec $\sim -35$d 50m. It extends towards SE out to $\sim15'$ ($\sim 85$ kpc) from the centre of the stellar body compared to $\sim7'$ ($\sim 40$ kpc) in the previous data.

The newly detected faint extension of the southern tail is located at velocities between $\sim1850$ and $\sim1970$ \kms, where no previous observations had detected \hi\ in this galaxy. Thanks to the MeerKAT data it is now apparent that, as we move farther from the stellar body towards SE, the \hi\ becomes more blue-shifted relative to the galaxy's systemic velocity $v_\mathrm{sys} = 2036$ \kms\ (Table \ref{table:n1427a}). The blue-shift of \hi\ in the tail exceeds the rotational blue-shift of \hi\ within the stellar body by $\sim100$ \kms.

\begin{figure*}
\includegraphics[width=17cm]{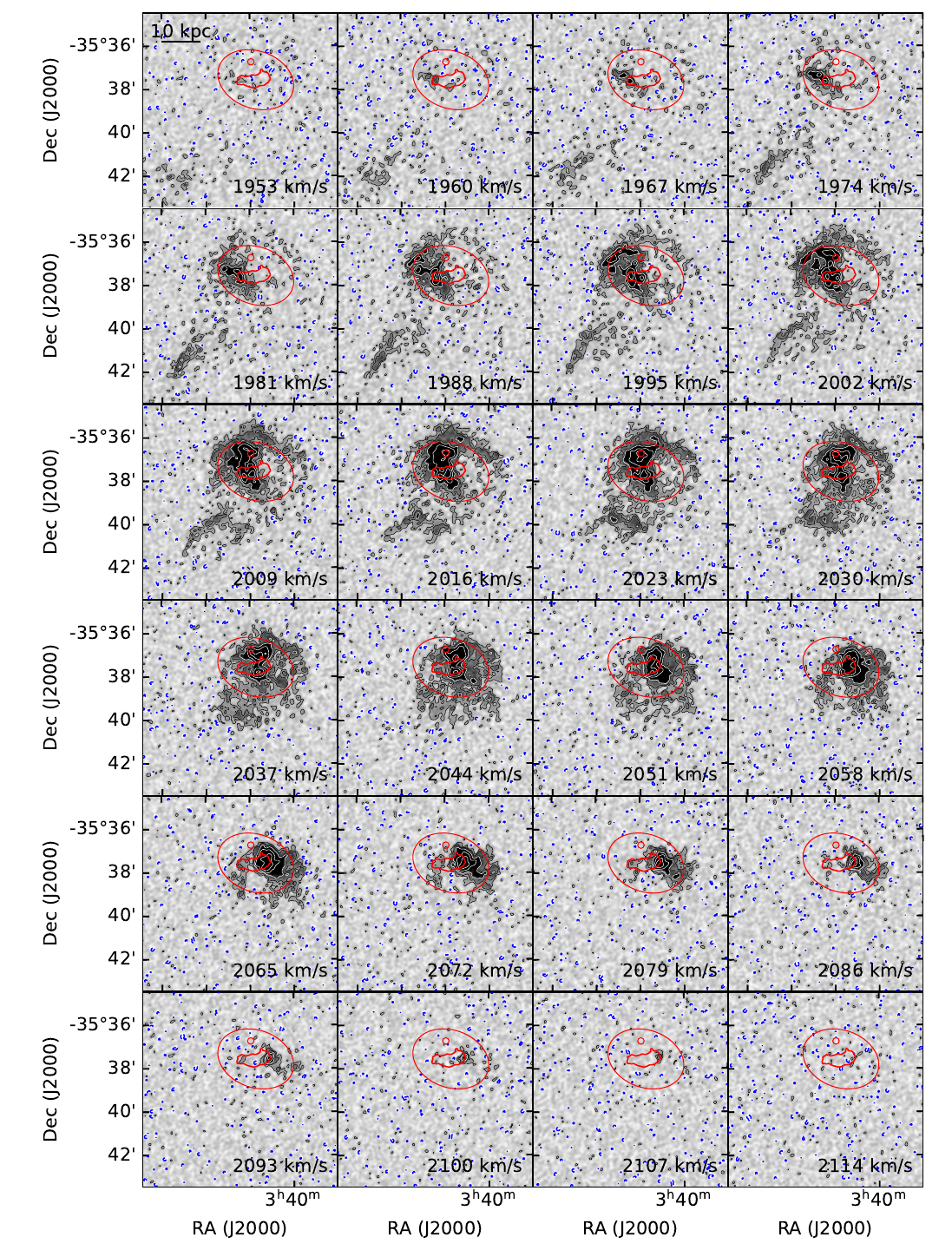}
\caption{Channel maps of the $11''$-resolution \hi\ cube of NGC~1427A. For visualisation purpose we binned the channels by a factor of 5 compared to the specs given in Sect. \ref{sec:data}, resulting in a channel width of 6.9 \kms. We use a linear grey scale from $-2\sigma$ to $+10\sigma$, where $\sigma$ is the noise level. The blue contour represents emission at $-2\sigma$. The black contours represent emission at 2 and $4\sigma$. The white contours represent emission at 8, 16, 32 and $64\sigma$. The red contours show the main features of the optical image (Fig. \ref{fig:opt}).}
\label{fig:t06chans}
\end{figure*}

\begin{figure*}
\centering
\includegraphics[width=9cm]{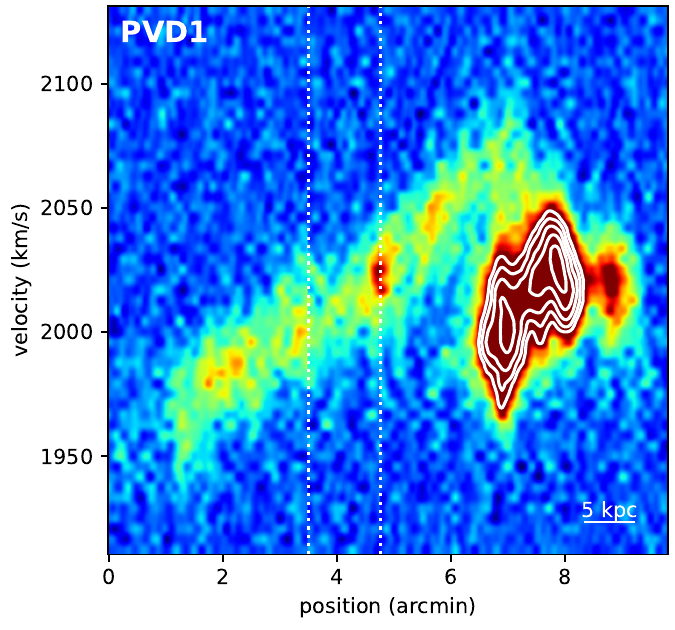}
\includegraphics[width=9cm]{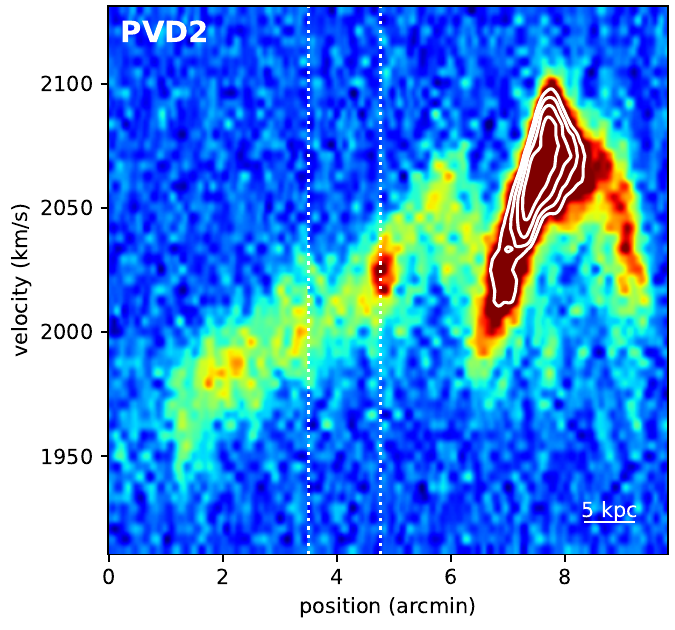}
\includegraphics[width=9cm]{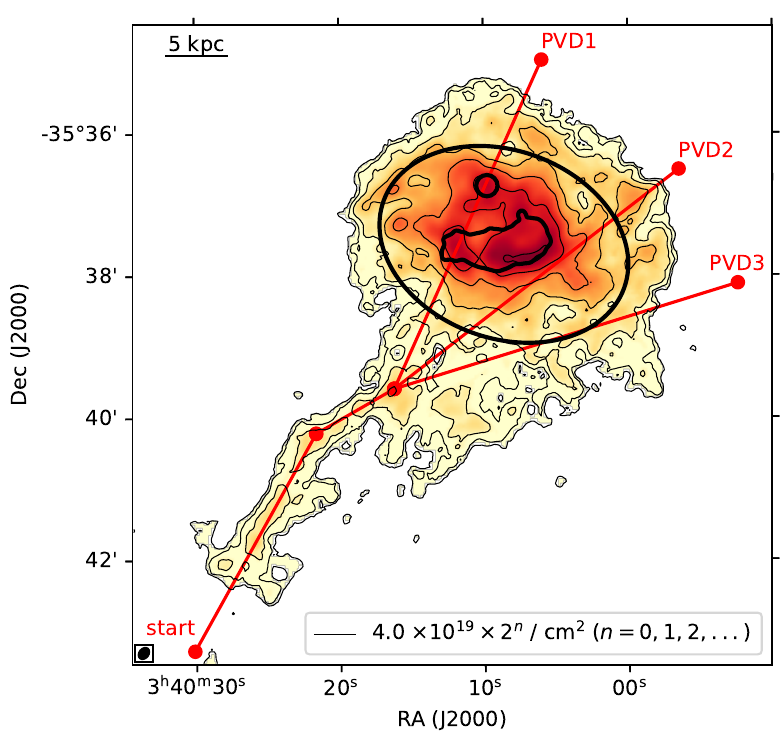}
\includegraphics[width=9cm]{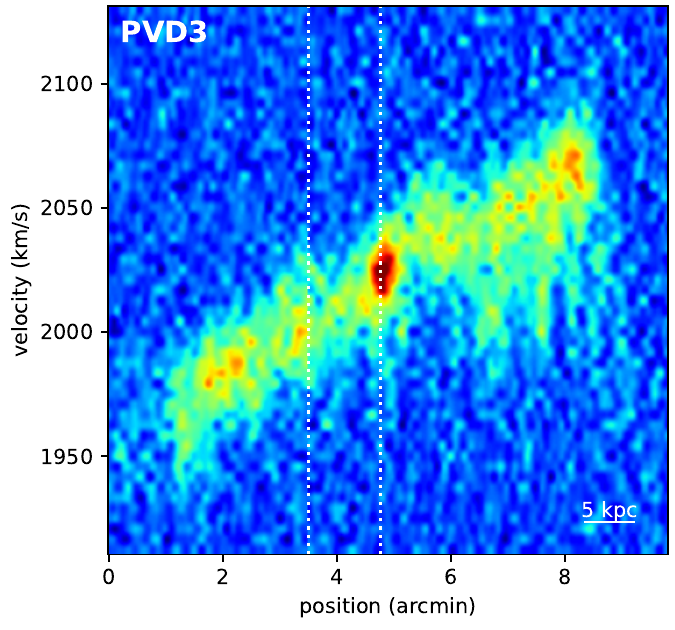}
\caption{\hi\ position-velocity diagrams along the three paths shown in red in the bottom-left panel. The diagrams are drawn using a width of $\sim20''$, twice the $11''$ angular resolution of the \hi\ cube used here. For visualisation purpose we binned the channels by a factor of 3 compared to the specs given in Sect. \ref{sec:data}, resulting in a channel width of 4.1 \kms. The three diagrams begin from the point labelled as ``start'' in the bottom-left panel and have a common first half, which consists of two straight segments. The end of these two segments is represented by the vertical dashed lines in the diagrams. The linear colour-scale of the diagrams ranges from $-2\sigma$ to $+8\sigma$, where $\sigma$ is the noise level. The white contours start at a $+10\sigma$ level and increase by a factor $\sqrt{2}$ at each step. The colour-scale and contours of the bottom-left panel are as in Fig. \ref{fig:tot11}.}
\label{fig:pvdtail}
\end{figure*}

\begin{figure*}
\centering
\includegraphics[width=18cm]{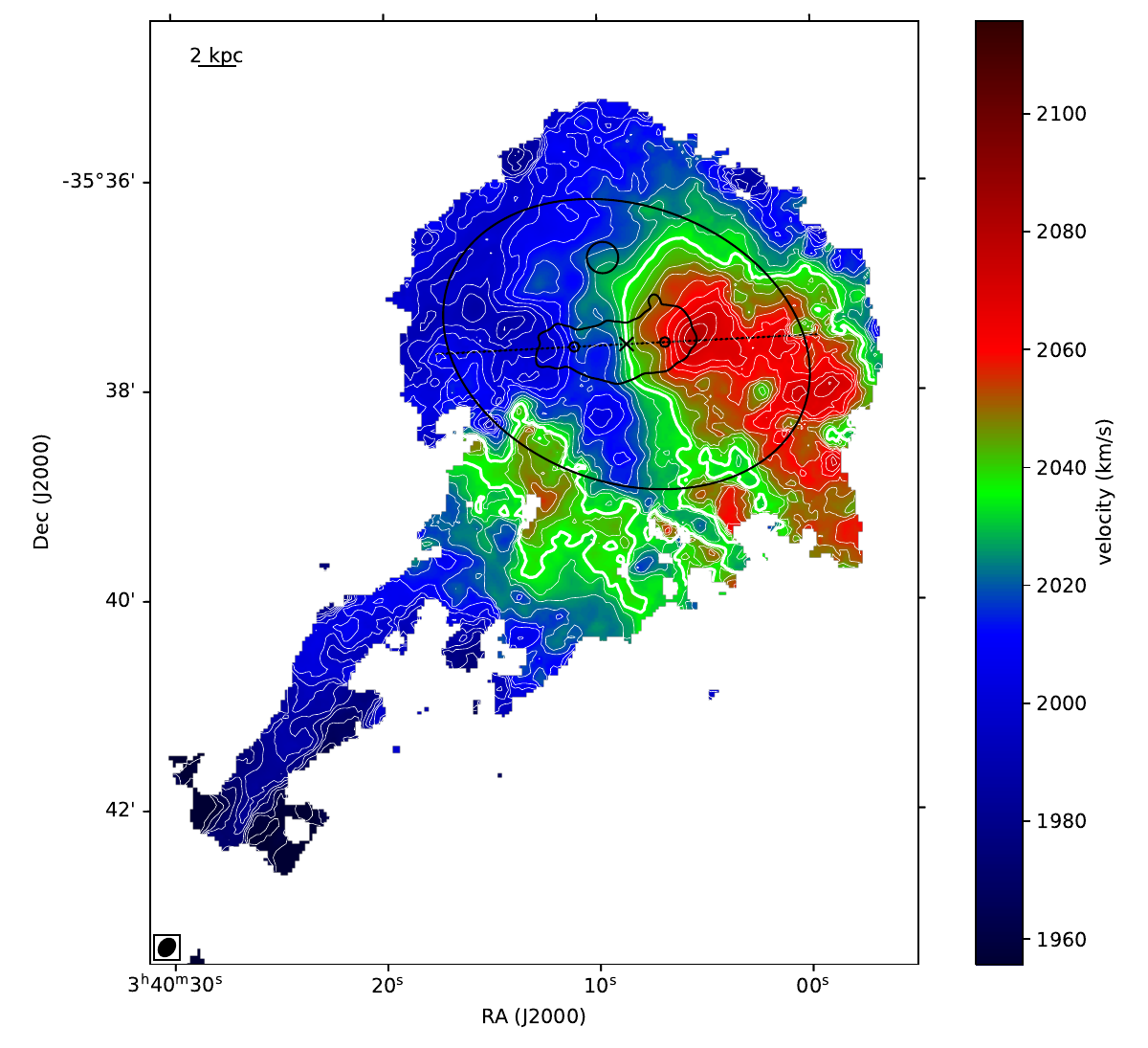}
\caption{Moment-1 velocity field at $11''$ resolution, showing the regular rotation within the bright stellar body of NGC~1427A, the PA warp at its edge, the complex velocity structure in the outer \hi\ envelope, and the velocity gradient along the densest part of the southern \hi\ tail. The isovelocity white contours are drawn at $\pm n\times 5$ \kms\ intervals ($n=0,1,2,...$) from the systemic velocity of 2036 \kms. The systemic velocity itself is highlighted with a thick white contour and is at the centre of the colourbar. The dynamical centre is marked with a black cross (see text). The black dotted line represents the axis along which we draw the position-velocity diagram shown in Fig. \ref{fig:pvdcen}. The two small, black open circles along that axis indicate the position of the velocity jumps visible in that diagram (see text). The black contours show the main features of the optical image (Fig. \ref{fig:opt}).}
\label{fig:vf11}
\end{figure*}

Further new insights come from the $11''$-resolution cube. In this case the \Nhi\ sensitivity is similar to that in \cite{leewaddell2018}, but at a $\sim7\times$ better angular resolution ($\sim42\times$ smaller beam area). Figure \ref{fig:t06chans} shows the channel maps of this cube, again binned by a factor of 5 in velocity to a channel width of 6.9 \kms\ for visualisation purpose. The southern \hi\ tail covers the same sky area and velocity range as in \cite{leewaddell2018}, i.e., we only see its brightest part, but we can now examine its structure in more detail. Moving from low to high velocity, \hi\ in this part of the tail moves steadily towards the stellar body approximately from SE to north-west (NW) up to $\sim2020$ \kms. This velocity gradient follows the one visible in the $66''$-resolution cube at larger radius and lower \Nhi, strengthening our result on the increasing blue-shift of \hi\ as a function of distance from the stellar body along the tail.

At even higher velocity the situation becomes more complex as the tail merges with the diffuse gas located at the southern edge of (and within) the stellar body. Here the tail's shape becomes more difficult to discern. One possibility is that the tail continues along an approximately east-west direction just outside the southern edge of the stellar body (the solid, red ellipse in the figure), reaching $\sim2040$ \kms\ and connecting with the western edge of the \hi\ distribution at even higher velocities. Another possibility is that the tail bends north and connects with the kinematically-anomalous \hi\ (i.e., gas not following the general rotation pattern of the main disc; see Sect. \ref{sec:dense}) located SE of the stellar body between $\sim2050$ and $\sim2080$ \kms. Interestingly, this anomalous gas reaches very close to the northern clump (red circle in the figure), although it does so at a significantly higher velocity ($\sim2070$ \kms) than the bright \hi\ most directly associated with the clump ($\sim2025$ \kms; see below).

We further illustrate the complex connection of the southern tail with the \hi\ in the stellar body in Fig. \ref{fig:pvdtail}. As shown in the bottom-left panel of this figure, we draw three position-velocity diagrams (PVD1, PVD2 and PVD3). All diagrams have a common first half, which follows the ``ridge'' of the tail along two straight segments. Then, three paths branch out. PVD1 goes through the eastern edge of the bright stellar body (at a position of $\sim7'$ along the horizontal axis) and the northern clump ($\sim8'$), following the anomalous gas discussed in the previous paragraph. This diagram confirms that the possible northern extension of the tail ``overshoots'' in velocity compared to the clump. However, at a position of $\sim7'$ the velocity gradient changes sign, and we find a faint distribution of \hi\ whose velocity decreases with increasing position, eventually connecting with the clump. Therefore, a connection between the southern tail and the northern clump along PVD1 cannot be ruled out, although the inversion of the velocity gradient would need to be explained.

Something similar can be seen in PVD2, which connects the tail with the opposite (western) edge of the bright stellar body. In this case the \hi\ continues along the same velocity gradient of the tail until $\sim6'$, after which the gradient changes sign and a faint gas distribution connects with the bright \hi\ in the stellar body. Again, if this connection is real the inversion of the velocity gradient needs to be explained.

Finally, PVD3 explores the possible continuation of the tail all the way to the western edge of the \hi\ distribution. Unlike in PVD1 and PVD2, in this case the \hi\ continues along the tail's velocity gradient until the end at $\sim8.5'$, where the gas velocity is as high as $\sim2080$ \kms. However, even in this case there appears to be a slight deviation from a simple increase of the velocity along the position axis, as the \hi\ velocity decreases briefly at $\sim6'$, and then starts increasing again. Another notable feature is the large \hi\ line width between $\sim6'$ and $\sim8'$, which could be inferred from Fig. \ref{fig:t06chans}, too.

Figure \ref{fig:hicontours} shows the full extent of the southern \hi\ tail. If, as in the original discovery paper by \cite{leewaddell2018}, we start measuring the tail's length from a distance of $3'$ from the stellar body --- i.e., at the outer optical isophote represented by the black dotted ellipse in the figure --- its length is now $\sim12'$  compared to the previous $\sim4'$. That is a total length of $\sim70$ kpc, $3\times$ longer than previously measured. If we include in this calculation the \hi\ clouds located past the tail's edge towards SE (Fig. \ref{fig:vfwide}), the length of the tail grows to $\sim50'$ or $\sim300$ kpc. The width of the \hi\ tail is remarkable, too: $\sim 7'$ in the $66''$-resolution image. Even making a generous correction for beam smearing, the intrinsic width must be at least $\sim5'$ or $\sim30$ kpc, comparable to the diameter of the main \hi\ body. The mass of the tail is $\sim0.4\times10^9$ \msun\ at low angular resolution (not including the scattered clouds, which would increase this value by $\sim5\%$). This is $\sim50\%$ more than in the original discovery paper and $\sim15\%$ of the total \hi\ mass of the galaxy.

\subsection{\hi\ in the stellar body}
\label{sec:dense}

Besides giving us a more detailed view of the southern tail, the $11''$-resolution channel maps in Fig. \ref{fig:t06chans} reveal the full complexity of the distribution and kinematics of the \hi\ associated with the stellar body. Overall, \hi\ within and at the boundary of the stellar body follows a clear velocity gradient, such that the line-of-sight velocity increases from east to west. However, the pattern is more complex than that of a simple rotating disc.

The bright \hi\ (in black in Fig. \ref{fig:t06chans}, with white contours overlaid) appears to include two components. The first is associated with the bright stellar body. It starts at its eastern edge at $\sim1960$ \kms, follows a regular velocity gradient as it moves westward within the bright optical isophote (PA $\sim100^\circ$; Sect. \ref{sec:intro}), and reaches its western edge at $\sim2100$ \kms. A slight \hi\ warp towards PA $\sim70^\circ$ is visible both at the eastern and western edge at a radius of $\sim 50''$ ($\sim5$ kpc). This PA twist of the \hi\ kinematical major axis is remarkably consistent with the PA twist observed in the optical isophotes (see outer red ellipse in the figure).

The regular \hi\ velocity pattern within the bright stellar body can be appreciated in Fig. \ref{fig:vf11}. Inside the bright optical isophote the velocity field resembles that of a rotating disc. Indeed, this figure allows us to identify an accurate dynamical centre and systemic velocity: RA = 03h 40m  8.7s, Dec = $-35$d 37m 34s, $v_\mathrm{sys} = 2036$ \kms\ (black cross in the figure). The velocity field shows a high level of symmetry about this point. Furthermore, the isovelocity contours around this point are perpendicular to the major axis and closely packed together, indicating the quick rise of the rotation curve around a central concentration of mass --- possibly the bright unresolved optical source visible in the central region of Fig. \ref{fig:opt}. The dynamical centre is $10''$ ($\sim1$ kpc) SW from the most common morphological centre given in the recent literature (e.g., 03h 40m 09.30s, $-35$d 37m 28.2s; \citealt{munozmateos2015,bai2015}). Just outside of the brightest optical isophote, the velocity field shows the aforementioned kinematical PA warp at a radius of $50''$ ($\sim 5$ kpc).

\begin{figure}
\centering
\includegraphics[width=7.2cm]{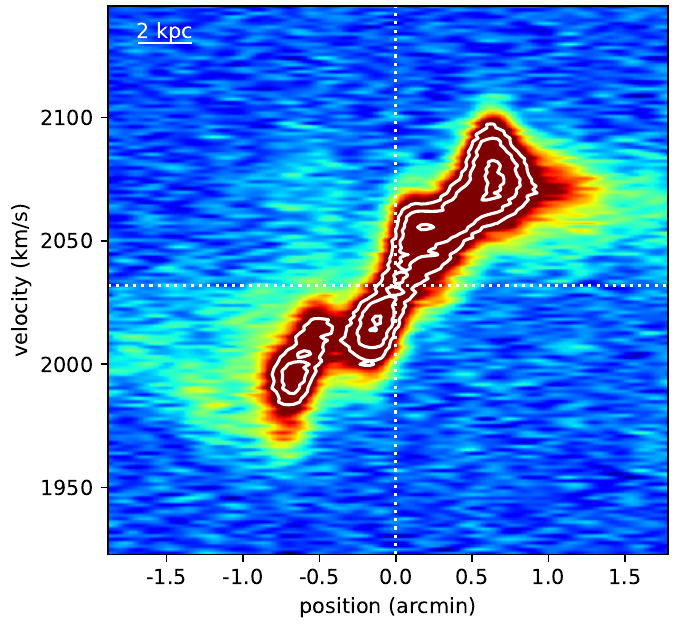}
\caption{\hi\ position-velocity diagram along the \hi\ kinematical major axis of NGC~1427A's bright stellar body constructed using the $11''$-resolution cube at full velocity resolution (1.4 \kms). The diagram is centred at the dynamical centre and systemic velocity given in the text (vertical and horizontal white lines, respectively), and the axis has PA $=93^\circ$. The linear colour-scale ranges from $-2\sigma$ to $+8\sigma$, where $\sigma$ is the noise level. The white contours start at a $+10\sigma$ level and increase by a factor $\sqrt{2}$ at each step. East is left, west is right.}
\label{fig:pvdcen}
\end{figure}

Despite the regular appearance of the inner velocity field in Fig. \ref{fig:vf11}, the kinematics of the \hi\ inside the bright stellar body exhibits some peculiarities, too. Figure \ref{fig:pvdcen} shows a position-velocity diagram along an axis passing through the dynamical centre and with PA $=93^\circ$ (perpendicular to the systemic isovelocity contour; see the black dotted line in Fig. \ref{fig:vf11}). Thanks to the excellent resolution of the data we reveal two clear velocity jumps on both sides of the galaxy at a radius of $\sim30''$ ($\sim3$ kpc; see the small, black open circles in Fig. \ref{fig:vf11}). This is inner to the $\sim 5$ kpc radius where the \hi\ kinematics appears to warp (see above) and inner to the radius where the optical PA changes (Sect. \ref{sec:intro}).

The second, bright \hi\ component in Fig. \ref{fig:t06chans} is centred on the northern clump at a velocity of $\sim2025$ \kms, consistent with the ionised-gas velocity measured at that location by \cite{chaname2000}. The direction of the \hi\ velocity gradient is similar to that in the bright stellar body: the velocity increases from east to west as the \hi\ moves along the northern plume (Sect. \ref{sec:intro}). In this case, however, the nature of the velocity gradient is less clear as we do not find convincing evidence of rotation about a well defined dynamical centre. In fact, while this \hi\ component stands out for its brightness, which is similar to that of \hi\ in the bright stellar body, its projected kinematics is fully consistent with that of the surrounding gas. A similar conclusion was reached by \cite{chaname2000} based on the kinematics of the ionised gas. The northern clump and the northern plume have a higher optical and \hi\ surface brightness than the surrounding regions (see also Fig. \ref{fig:tot11}) but, kinematically, they do not appear as a distinct system.

Fainter gas (in grey in Fig. \ref{fig:t06chans},  with overlaid black contours) is found at, and slightly beyond, the edge of the stellar body (red ellipse in the figure). As shown by the \Nhi\ $\sim10^{20}$ cm$^{-2}$ contours in Figs. \ref{fig:hicontours} and \ref{fig:tot11}, this outer \hi\ is aligned with the outer optical isophote: it has the same elliptical shape, PA and ellipticity, and only a slightly larger semi-major axis. In the NE and SW, the outer \hi\ extends the kinematical PA warp started at a radius of $\sim50''$ (Fig. \ref{fig:vf11}). In the SE, as discussed in Sect. \ref{sec:tails}, the outer \hi\ visible between $\sim2020$ and $\sim2050$ \kms\ merges with the start of the southern \hi\ tail without a clear boundary between them. In the same region, the kinematically anomalous \hi\ stretching towards the northern clump at velocities between $\sim2050$ and $\sim2080$ \kms\ may be an extension of the tail. Finally, in the NW, just outside the red ellipse representing the outer optical isophote, we find an \hi\ ``arm'' at velocities $\sim1990$ to $\sim2040$ \kms\ with only a weak velocity gradient (the line-of-sight velocity increases weakly from east to west). This arm is clear in the channel maps and appears as the main deviation between \hi\ and outer optical isophotes in Fig. \ref{fig:tot11}.

The kinematics of the faint, outer \hi\ is complex and does not completely follow that of a rotating disc. This is somewhat surprising considering that the morphology of the outer \hi\ in Fig. \ref{fig:hicontours} is regular and traces well the elliptical, outer optical isophote. Yet, the velocity field shown in Fig. \ref{fig:vf11} leaves no doubts: while the \hi\ kinematical major axis describes an S shape, in line with the PA twist of the optical isophotes, the \hi\ kinematical minor axis (i.e., the thicker, white $v_\mathrm{sys}$ isovelocity contour overlaid on bright green pixels in the figure) does not follow a corresponding S shape, as it should in case of a simple PA warp. Instead, it follows a U shape, where the ``U'' is sideways, opening to the west: most of the \hi\ is found at blue-shifted velocities (``outside'' the U), and only a small SW wedge is found at red-shifted velocities (``inside'' the U). Furthermore, the maximum offset from $v_\mathrm{sys}$ is $\sim10$ \kms\ larger on the approaching side than on the receding side of the outer disc. Finally, \hi\ in the aforementioned NW arm is mostly at blue-shifted velocities.

\begin{figure}
\centering
\includegraphics[width=9cm]{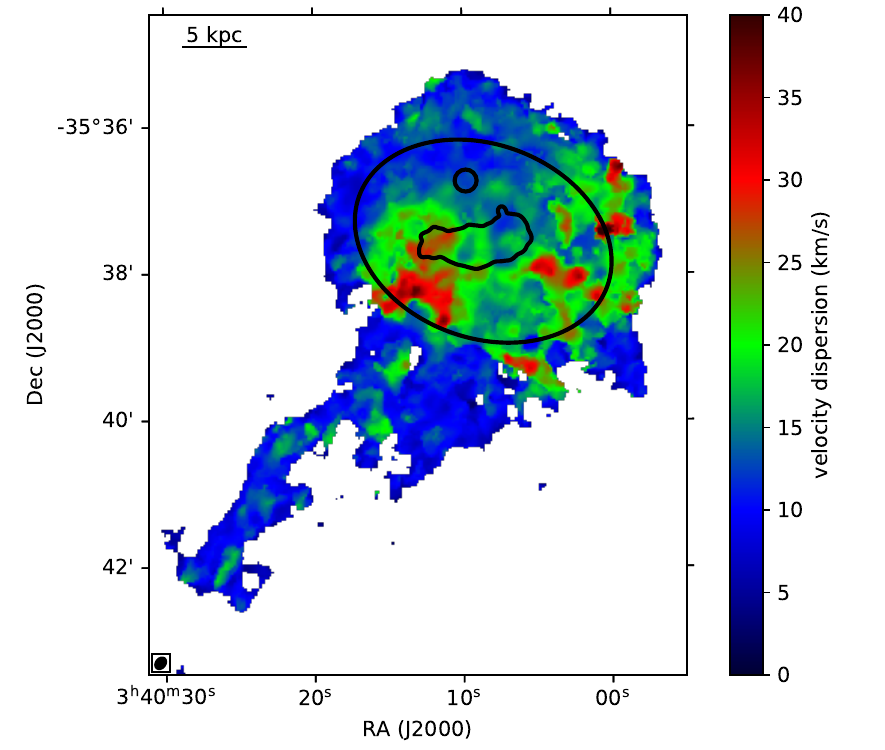}
\caption{Moment-2 velocity dispersion field at $11''$ resolution. The black contours show the main features of the optical image (Fig. \ref{fig:opt}).}
\label{fig:vdisp11}
\end{figure}

The complex \hi\ kinematics is confirmed by the velocity dispersion field in Fig. \ref{fig:vdisp11}. Despite the $\sim1$ kpc resolution and, therefore, the limited beam smearing, most of the \hi\ within and immediately around the stellar body is characterised by broad line profiles with dispersion between $\sim20$ and $\sim40$ \kms, a few times larger than the typical value of $\sim10$ \kms\ for the resolved \hi\ line. This is largely a result of the presence of anomalous gas across most of the \hi\ body. We have already mentioned the anomalous component SE of the bright stellar body, but Fig. \ref{fig:t06chans} shows anomalous gas in the west, too. There, within the outer optical isophote, \hi\ emission is visible at velocities between $\sim1980$ and $\sim2030$ \kms, which is lower than systemic even though that is the receding side of the disc, resulting in a high velocity dispersion. Only in the NE quadrant (in particular at the location of the northern clump and the northern plume) as well as along the southern tail do we find more common \hi\ velocity dispersion values of $\sim10$ \kms.

In summary, \hi\ within and around the edge of the stellar body of NGC~1427A makes up $\sim85\%$ of the detected \hi\ mass: $\sim1.9\times10^9$ \msun\ out of a total of $\sim2.3\times10^9$ \msun\ (the remaining $\sim0.4\times10^9$ \msun\ is found in the southern tail). To the extent that the aforementioned components can be separated in the data cube, this \hi\ mass includes: $\sim0.8\times10^9$ \msun\ of regularly rotating gas associated with the inner, bright part of the stellar body; $\sim0.4\times10^9$ \msun\ associated with the northern clump and the northern plume, exhibiting a clear velocity gradient but neither obvious signs of rotation nor anomalous kinematics compared to the surrounding gas, such that it might not be a distinct system; and $\sim0.7\times10^9$ \msun\ distributed in an elliptical outer region characterised by complex kinematics (an S-shape warp of the kinematical major axis but a U-shape warp of the kinematical minor axis, and plenty of anomalous gas which results in an unusually high velocity dispersion) and by an excess of \hi\ along a blue-shifted NW arm. In the next Section we attempt to explain all these results in a consistent way.

\begin{figure*}
\centering
\includegraphics[width=9cm]{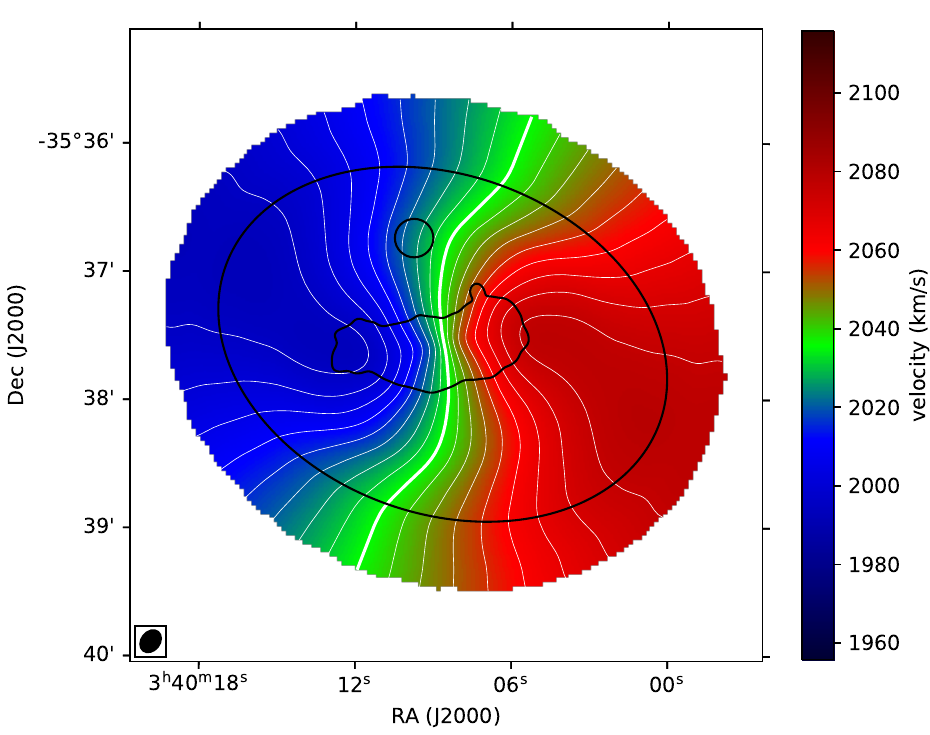}
\includegraphics[width=9cm]{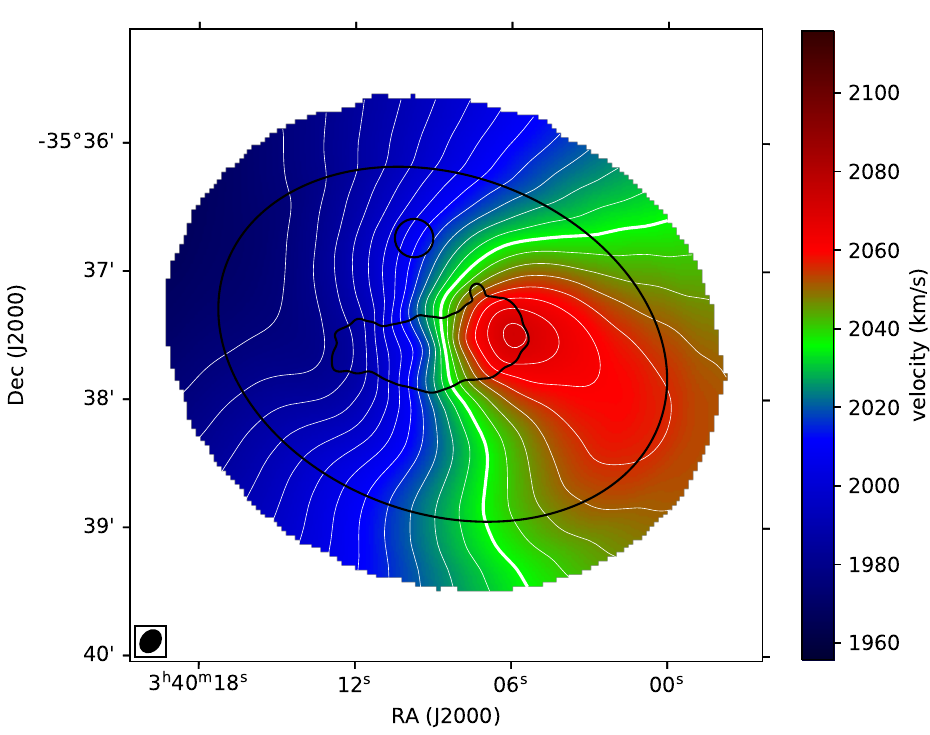}
\caption{\it Left panel. \rm Velocity field of the \hi\ model cube built to match the observed S-shape of the kinematical major axis (Fig. \ref{fig:vf11}). \it Right panel. \rm The same model as in the left panel, but now allowing for a variation of $v_\mathrm{sys}$ with radius in order to match the observed U-shape of the kinematical minor axis, too (Fig. \ref{fig:vf11}). We interpret the better match of the latter model to the data in terms of ram pressure changing the outer gas kinematics.}
\label{fig:tirific}
\end{figure*}

\section{Ram pressure stripping and tidal interactions in NGC~1427A}
\label{sec:interpr}

The new observation of NGC~1427A taken as part of the MeerKAT Fornax Survey reveals a remarkably complex distribution of \hi. As discussed in Sect. \ref{sec:intro}, having established that ram pressure is shaping the long, southern \hi\ tail \citep{serra2023}, the main open questions concern how deep into the stellar body ram pressure affects the ISM, whether and what type of tidal forces are acting on the galaxy, and what the sequence of --- and interplay between --- hydrodynamical and tidal interactions might be. In this Section we attempt to answer these questions based on the results described in Sect. \ref{sec:results}.

\subsection{Ram pressure}
\label{sec:rps}

We find two signatures of ram pressure. The first is related to the one-sided, starless, southern \hi\ tail. The tail was originally discovered by \cite{leewaddell2018} but, thanks to the much improved sensitivity and angular resolution of our MeerKAT data \citep{serra2023}, we can now study it in a more complete and detailed way. The tail extends out to 70 kpc from the edge of the stellar body towards SE. The \hi\ recessional velocity decreases with increasing radius across the full length of the tail, reaching values lower than those in the stellar body by $\sim100$ \kms. The direction of the velocity gradient along the \hi\ tail could be readily explained in terms of a blue-shifting ram pressure given that NGC~1427A is strongly red-shifted relative to the ICM of Fornax (Sect. \ref{sec:intro}). Therefore, our interpretation is that, due to the motion of NGC~1427A through the ICM, the tail and, in fact, the galaxy as a whole experience a ram-pressure wind directed towards us along the line of sight and towards SE on the plane of the sky.

Following the direction and velocity gradient of the tail past its end, we detect a distribution of scattered, starless \hi\ clouds, even more blue-shifted than gas in the tail (by up to $\sim100$ \kms), each with a mass between $\sim10^6$ and $\sim3\times10^6$ \msun, and reaching as far as 300 kpc from the galaxy. These clouds could be the extreme end of the tail, the most ancient remnant of the passage of NGC~1427A through the ICM. Many of their properties, such as their position, mass, velocity, as well as the gap between the clouds and the main tail, are qualitatively consistent with the mixing-based ram-pressure stripping model by \cite{tonnesen2021}. This suggests that these distant clouds might consist of a mix of \hi\ stripped from the galaxy and gas that has cooled from the ICM.

\cite{leewaddell2018} suggested that the southern \hi\ tail might have started tidal and, therefore, might be one of two tidal tails. We confirm this possibility and discuss it in detail in Sect. \ref{sec:tidal}. However, even in the case of a tidal origin, subsequent events have caused the southern tail alone to stretch to its considerable length towards SE and to develop the observed velocity gradient, thus implying that the galaxy is experiencing a ram-pressure wind with the orientation discussed above. The large width of the tail ($\sim30$ kpc) is also qualitatively consistent with hydrodynamical effects being at play, as normally tidal tails are much narrower \citep[e.g.,][]{appleton1987,rots1990,hibbard1994,hibbard2001b,clemens1999,langston2001,mundell2004}. Within this framework, the twin tidal tail would necessarily be located in the NW, where the same ram pressure would push it to lower velocities and back towards the stellar body. In Sect. \ref{sec:tidal} we discuss whether the NW \hi\ arm located just outside the outer optical isohpote (Figs. \ref{fig:tot11} and \ref{fig:t06chans}) might fit into this description.

The second signature of ram pressure is the U shape of the kinematical minor axis of the \hi\ velocity field in Fig. \ref{fig:vf11}. This shape is inconsistent with the PA warp from $\sim100^\circ$ to $\sim70^\circ$ revealed both by the S-shaped \hi\ kinematical major axis and by the PA twist of the optical isophotes (Sect. \ref{sec:intro}). Yet, despite the complexity of the \hi\ channel maps --- partly a result of the presence of kinematically anomalous gas --- the fairly regular appearance of the \hi\ velocity field suggests that the gas motions are sufficiently ordered, and that a relatively simple interpretation of the data should be possible. Indeed, the same blue-shifting ram pressure which explains the southern \hi\ tail's velocity gradient could in principle explain the U-shaped kinematical minor axis, too (e.g., \citealt{haan2014}). We explore this with the aid of \hi\ model data cubes built with \texttt{TiRiFiC} \citep{jozsa2007}.

We assume that the \hi\ rotates on a disc with the dynamical centre given in Table \ref{table:n1427a} and Sect. \ref{sec:dense}. We let \texttt{TiRiFiC} solve for a constant inclination, a constant PA and a rotational velocity dependent on radius out to $50''$ from the centre, while we keep the rotation velocity constant farther out and let the PA (and the inclination) vary with radius to accomodate the twist of the \hi\ kinematical major axis (we perform some of this initial fitting with Fully Automated TiRiFiC; \citealt{kamphuis2015}). Since large parts of the \hi\ distribution are inconsistent with such a simple model (e.g., the U-shaped kinematical minor axis, the southern tail and the numerous clouds of anomalous gas), we take the best-fitting model as a starting point and adjust some of the model parameters manually to build an \hi\ cube which, based on visual inspection, reproduces the following main features of the data: the observed velocity range, the ellipticity and the shape of the kinematical major axis. We show the resulting model velocity field in Fig. \ref{fig:tirific} (left panel). The PA of the model changes from $\sim100^\circ$ in the inner region to $\sim80^\circ$ further out, the inclination from $\sim55^\circ$ to $\sim25^\circ$, and the rotational velocity reaches $\sim100$ \kms, in excellent agreement with the baryonic Tully-Fisher relation \citep[e.g.,][]{lelli2019} given the measured baryonic mass of $4.6\times10^9$ \msun\ (Table \ref{table:n1427a}). As anticipated, a simple PA warp cannot reproduce the observed U shape of the kinematical minor axis.

Next, we modify our simple model by letting \texttt{TiRiFiC} solve for a varying $v_\mathrm{sys}$ as a function of radius beyond the inner $50''$. This change is meant to represent an additional, radius-dependent velocity component along the line of sight, which would be caused by the blue-shifting ram pressure discussed above. Again, we adjust the model parameters manually to maximise the visual match with the aforementioned main features of the data, but now matching also the shape of the kinematical minor axis. We show the resulting model velocity field in Fig. \ref{fig:tirific} (right panel). The match is reasonably good when $v_\mathrm{sys}$ decreases monotonically from 2036 \kms\ at a radius of $50''$ to $\sim2005$ \kms\ at a radius of $3'$. This model also matches the observed kinematical lopsidedness along the kinematical major axis, such that the maximum offset from $v_\mathrm{sys}$ is larger on the approaching side of the galaxy compared to the receding side (Sect. \ref{sec:dense}). We consider further tuning of the model unnecessary given the complexity of the data.

The increasing shift of $v_\mathrm{sys}$ with radius of our model is consistent with a blue-shifting ram pressure whose effect is stronger at larger radius because of the lower restoring force. Ram pressure can thus explain at once both the velocity gradient in the southern \hi\ tail and the peculiar appearance of the \hi\ velocity field in Fig. \ref{fig:vf11}.

Ram pressure had already been proposed by \cite{chaname2000}, \cite{sivanandam2014} and \cite{mora2015} to explain the star-forming front at the SW edge of NGC~1427A's bright stellar body (Sect. \ref{sec:intro} and Fig. \ref{fig:opt}). This proposal was challenged by \cite{leewaddell2018} on the ground that the required ram pressure should have caused an \hi\ tail directed towards NE and not, as observed, towards SE. Our new, high-resolution data bring additional evidence against ram pressure pushing gas towards NE, as Fig. \ref{fig:tot11} shows no signs of compression of the \hi\ distribution on the SW side of the galaxy. We see no such compression at the location of the star-forming front, which is embedded deep into the \hi\ distribution, nor do we see a SW compression at larger radius, where the \hi\ could have been displaced more easily. Based on these results, we conclude that the intense activity in the SW star-forming front must be caused by some other process.

A related point is that we find no indication of a compression of the outer \hi\ distribution in any direction --- not even on the NW side opposite to the southern \hi\ tail. This suggests that the SE-bound component of the ram-pressure wind generated by the motion of NGC~1427A through the ICM is too weak to significantly displace in RA and Dec gas at the edge of (or well within) the stellar body. A further implication is that the southern \hi\ tail must be made of gas that was pulled outside of the stellar body by some other process (e.g., a tidal interaction, with an additional possible contribution by ram pressure along the line of sight). Once there, due to its weak gravitational binding to the galaxy, the gas could be pushed further towards SE by this weak component of the ram-pressure wind. The direction of this displacement points almost exactly away from Fornax' central galaxy NGC~1399 on the plane of the sky and, therefore, suggests that NGC~1427A is currently approaching the centre of the cluster in projection.

Based on our data, we conclude that the main component of the galaxy motion through the ICM and, therefore, of the ram-pressure wind felt by the galaxy because of this motion, is the one along the line of sight. This blue-shifting component of the ram-pressure wind is sufficiently strong not only to create the velocity gradient along the southern tail, but also to transfer enough momentum to \hi\ within the stellar body and bend the kinematical minor axis into a U shape. At a more speculative level, it is possible that the main \hi\ body has a slight bowl shape as a result of this hydrodynamical push along the line of sight, whose effect increases with radius. Furthermore, since the ram-pressure wind is stronger along the line of sight than on the plane of the sky it is likely that the southern tail has a significant extension towards us. Its real length might be much more than the projected value of $\sim70$ kpc. Finally, the only gas apparently unaffected by ram pressure is that located within the high-surface-brightness region of the stellar body (radius $\lesssim 50''$ or 5 kpc, which is approximately the half-light radius). Presumably, the restoring force in this region is sufficient to resist ram pressure and, therefore, the gas kinematics remains regular (Fig. \ref{fig:vf11}). This is consistent with our comparison between the strength of the ram pressure and that of the average restoring force presented in Sect. \ref{sec:intro}.

The ultimate demonstration that a blue-shifting ram pressure is shaping the \hi\ kinematics not only in the southern tail but also within the stellar body would be to show that, in the latter region, the stellar kinematics does not follow the \hi\ kinematics. That is, we expect only the \hi\ to show a U-shaped kinematical minor axis caused by ram pressure, while the stars should exhibit an S-shaped minor axis as well as major axis, in line with the optical PA twist. In order to test this prediction we would need to study the stellar kinematics out to large radius but, to the best of our knowledge, no such data are currently available.

\subsection{Tidal interaction}
\label{sec:tidal}

As argued by \cite{leewaddell2018} based on the discovery of the southern \hi\ tail and confirmed in Sect. \ref{sec:rps} from the lack of a compression of the outer \hi\ disc, ram pressure cannot be responsible for the SW star-forming front in NGC~1427A (Fig. \ref{fig:opt}). What is its cause then? More generally, what is the cause of the irregular morphology of this galaxy, of its asymmetric inner region, of the shifting centre of the isophotes with radius, and of its optical PA twist?

An important new result of our high-resolution MeerKAT data is the close correspondence between the distribution of stars and \hi\ in this galaxy (with a few notable exceptions, such as the southern tail). In particular, the densest \hi\ follows closely the distribution of high-surface-brightness optical light at the location of both the bright stellar body and the northern plume centred on the northern clump (Fig. \ref{fig:tot11}). Furthermore, the outer \hi\ contours have the same elliptical shape of the outer optical isophotes (Fig. \ref{fig:tot11}), and the PA warp of the \hi\ kinematical major axis follows closely the PA twist of the optical isophotes (Fig. \ref{fig:vf11}). Stars and \hi\ track one another across most of this system. This implies that the forces at play affect equally the stars and the ISM and, therefore, must be tidal in nature.

A further argument in favour of a tidal interaction is the complex connection of the southern \hi\ tail with the main \hi\ body. As discussed in Sect. \ref{sec:tails} and illustrated by Fig. \ref{fig:pvdtail}, we cannot find a path which, following the velocity gradient induced along the tail by ram pressure (Sect. \ref{sec:rps}), smoothly connects gas in the tail to gas in the main body. Instead, we always find an inversion of the velocity gradient in the regions closer the stellar body (see in particular PVD1 and PVD2 in Fig. \ref{fig:pvdtail}). That is, in the initial segment of the tail (whichever it might be; see the discussion in Sect. \ref{sec:tails}) the gas velocity actually increases with increasing distance from the centre of the galaxy, opposite to the trend along the rest of the tail. Therefore, in that initial segment, the \hi\ is not being stripped from the main body by the blue-shifting ram pressure. A tidal interaction appears to be the most natural alternative explanation for the existence of the tail, and it is only after this initial, tidal segment that the velocity gradient changes sign and the gas starts to shift towards progressively lower velocities with increasing radius as a result of ram pressure. Therefore, having established above that NGC~1427A experiences a SE-bound, blue-shifting ram-pressure wind because of its motion through the ICM, we now argue that tidal forces must be at play, too.

Based on a set of hydrodynamical simulations, \cite{mastropietro2021} proposed that NGC~1427A is being distorted by a tidal interaction with the overall gravitational potential of the Fornax cluster centred on the galaxy NGC~1399. In particular, close to pericentre, the tidal forces of a Fornax-like cluster potential would be sufficient to disturb the stellar body (and the ISM) of a small satellite galaxy. In their interpretation, these forces would be responsible for the optical PA twist and elongation of the outer stellar body (Fig. \ref{fig:opt}) --- and we can now add the \hi\ kinematical PA warp to this list.

We find that the proposal by \cite{mastropietro2021} has one significant limitation: NGC~1427A might be too massive for the tidal field of Fornax to cause the observed distortions. In order to show this we compare the $50''$ radius ($\sim5$ kpc) where the shape of the inner, bright stellar body starts to change into the outer elliptical one, to a generous (i.e., as small as reasonable) estimate of the tidal radius $r_\mathrm{t}$, beyond which the tidal forces of the cluster potential exceed the binding forces of the galaxy. Under the assumptions of circular orbit and of a point-like host mass $M$ (which minimise $r_\mathrm{t}$; see \citealt{gajda2016}), a satellite with mass $m$ at distance $R$ from the host has $r_\mathrm{t} = [m/M / (3+m/M)]^{1/3} \times R$ \citep{binney1987}. The tidal radius decreases with decreasing $R$ and $m/M$. Therefore, we minimise our estimate on $r_\mathrm{t}$ by taking $R\sim130$ kpc, which is the projected- and, therefore, minimum current distance of NGC~1427A from NGC~1399 (Sect. \ref{sec:intro}). The cluster's dynamical mass within this radius is $M\sim10^{13}$ \msun\ \citep{drinkwater2001a}. The mass of NGC~1427A can be estimated by scaling the sum of the stellar and \hi\ mass ($\sim5\times10^9$ \msun; Table \ref{table:n1427a}) by an appropriate baryon fraction. To minimise  $r_\mathrm{t}$ we take a baryon fraction of 5\% (higher than typical estimates for small galaxies) and thus obtain $m\sim10^{11}$ \msun\ and $m/M\sim10^{-2}$. The result is $r_\mathrm{t}\sim20$ kpc. Alternatively, we could take the dynamical mass enclosed within the last point of the kinematical model described in Sect. \ref{sec:rps}. This would give $m\sim5\times10^{10}$ \msun\ (baryon fraction $\sim10\%$), $m/M\sim5\times10^{-3}$ and $r_\mathrm{t}\sim15$ kpc. Both estimates of $r_\mathrm{t}$ are significantly larger than the $\sim5$ kpc radius where the stellar body changes orientation. In fact, they are well outside the outer isophote of the stellar body in Fig. \ref{fig:opt} and the outer \hi\ contour in Fig. \ref{fig:tot11}. In order to have $r_\mathrm{t}=5$ kpc we would need a 25 to 50 times smaller mass ratio, $m/M = 2\times10^{-4}$. This would correspond to $m=2\times10^9$ \msun, which is just about the \hi\ mass of NGC~1427A alone and would thus require a baryon fraction of 100\%.

This calculation shows that, for a galaxy with the mass and current position of NGC~1427A within Fornax, the tidal forces associated with the cluster gravitational potential are not sufficiently strong to cause the observed distortion of the stellar body at a radius of $\sim5$ kpc. Indeed, all satellite galaxies matching the disturbed morphology of NGC~1427A in the simulations of \cite{mastropietro2021} are significantly smaller: the most massive has \mstar\ $\sim6\times10^8$ \msun, 4 times less massive than NGC~1427A; while the satellite whose distortion and tails are shown in that paper (e.g., their Figure 3) has \mstar\ $\sim9\times10^7$ \msun, 30 times less massive than NGC~1427A. Therefore, independent of our estimates of $r_\mathrm{t}$ above, the simulations of \cite{mastropietro2021} do not seem to apply straightforwardly to our target. 

A possible way to reconcile those simulations with the case of NGC~1427A is if the galaxy has already passed pericentre, i.e., was once much closer to NGC~1399 and is now on the way out of the cluster. Since $r_\mathrm{t}$ is directly proportional to the distance $R$, a pericentric distance of $\sim20$ kpc from NGC~1399 would bring the tidal radius estimate closer to the required $\sim5$ kpc. Two problems arise though. First, it is difficult to imagine that the extended \hi\ disc of NGC~1427A (radius $\gtrsim10$ kpc) would survive such a close encounter with NGC~1399 maintaining its outer regions' elliptical shape intact, especially considering that its speed relative to the ICM would have been even larger than the current $\sim600$ \kms\ (Sect. \ref{sec:intro}). Second, if the galaxy is now past pericentre (but still necessarily closeby given its high speed) we would expect a tail of ram-pressure stripped \hi\ directed towards the cluster centre, i.e., towards NW, opposite to the observed one. An indirect confirmation of the latter issue is the fact that most snapshots matching the properties of NGC~1427A in the simulations by \cite{mastropietro2021} show satellite galaxies before pericentric passage.

We thus find that, while certainly relevant to the evolution of satellite galaxies in clusters, the conclusions of \cite{mastropietro2021} are not applicable to NGC~1427A. Therefore, while we argued above that the galaxy must be subject to tidal forces, we conclude that these forces cannot be due to an interaction with the cluster gravitational potential. The only option left is that of a recent or ongoing tidal interaction --- possibly a merger --- between two galaxies.

We start our discussion of a possible galaxy-galaxy interaction by estimating its timescale. Given our idea that the southern \hi\ tail started tidal and was later stretched by ram pressure, we can estimate the time passed since the beginning of the tidal interaction by calculating the time required for the tail to reach its current projected length of 70 kpc. Assuming that the speed of NGC~1427A through the ICM on the plane of the sky is $\sim200$ \kms, i.e., $\sim1/3$ of the value measured along the line of sight (qualitatively motivated by the discussion in Sect. \ref{sec:rps}), the tidal interaction must have started at least $\sim300$ Myr ago. This is a lower limit because: \it i) \rm it takes time to accelerate the stripped gas into the ICM \citep[e.g.,][]{tonnesen2021}; and \it ii) \rm the tail is $\sim4$ times longer if we include the SE \hi\ clouds of Fig. \ref{fig:vfwide}.

A timescale of $\sim300$ Myr is significantly longer than the $\sim4$ Myr age of the SW star-forming regions used in previous work to date the tidal interaction \citep[e.g.,][]{leewaddell2018}. It is sufficiently long to allow for a high-speed encounter between NGC~1427A and another cluster member, which could have then drifted away by as much as a few 100's kpc. For example, for a fly-by occurred 300 Myr ago at a speed of 1000 \kms\ ($\sim3\times$ the velocity dispersion of Fornax; \citealt{maddox2019}) the two interacting galaxies would now be 300 kpc apart. This is the approximate projected distance between NGC~1427A and NGC~1437A (towards SE; line-of-sight velocity difference $\sim1150$ \kms) or ESO~358-51 (towards north; line-of-sight velocity difference $\sim300$ \kms). Both galaxies have a disturbed stellar body and a one-sided \hi\ tail \citep{serra2023}, and could therefore be the interacting companion. A high speed encounter occurred a few 100's Myr ago is, therefore, a viable option, but a set of tailored simulations might be required to understand under what circumstances it would unbind enough \hi\ to explain our observations.

The other possibility is that of an on-going, low-speed interaction resulting in a merger, as discussed in most papers on NGC~1427A  ever since \cite{cellone1997}. Chance encounters with a low relative speed are unlikely in a cluster and, therefore, it is probable that the two merging galaxies fell into Fornax as a pair and continued interacting on their way in. For relative speeds of the order of $\sim100$ \kms, the above timescale  of $\sim300$ Myr would require the two merging galaxies to be within $\sim30$ kpc of one another. As discussed by \cite{leewaddell2018} and confirmed in Sect. \ref{sec:results}, the field around NGC~1427A hosts no such nearby galaxy with a sufficiently large mass or signatures of a past interaction to be a plausible merging companion. Rather, given the complexity of the optical morphology and of the \hi\ distribution and kinematics of this galaxy, we consider it more likely that the merger is in a sufficiently advanced stage that the two systems are already in the process of settling within the remnant.

Given the optical appearance of the galaxy, it is natural to consider the possibility that the northern clump is the remnant of a merging satellite now largely disrupted. Part of the satellite would still be visible in the form of the northern plume centred on the northern clump. The main objection to this hypothesis is that the kinematics of gas in the northern plume and the northern clump is consistent with that of gas in adjacent regions and, therefore, the northern clump does not appear to be an intruder within NGC~1427A (modulo projection effects). This was pointed out by \cite{chaname2000} based on optical long-slit observations of the ionised gas and is confirmed by our mapping of the distribution and kinematics of \hi\ across the full system at $\sim1$ kpc and 1.4 \kms\ resolution. In the rest of this section we discuss the merger hypothesis in more detail, but the \cite{chaname2000} objection still stands.

One consequence of the merger would be the creation of two tidal tails made of matter originally distributed at the outskirts of the main progenitor --- possibly mostly \hi. One could be the southern \hi\ tail, the other the NW \hi\ arm discussed in Sect. \ref{sec:dense} and most evident in the channels maps of Fig. \ref{fig:t06chans} at velocities $\sim1990$ to $\sim2040$ \kms\ just outside the outer optical isohpote. Normally, the two tails would be relatively similar in size and shape. However, as argued in Sect. \ref{sec:rps}, once outside the stellar body matter in both tails would be exposed to ram pressure, with only a weak gravitational binding to the galaxy. The southern tail would be stretched towards SE and would develop the observed velocity gradient. The northern tail would be pushed back towards the stellar body and, again, be subject to an overall shift towards low recessional velocities. The latter would be consistent with the blue-shift of the NW \hi\ arm discussed in Sect. \ref{sec:dense}, which we were not able to model (Fig. \ref{fig:tirific}). Ram pressure could thus explain the different shape of the two tails despite their common origin. We note that their \hi\ masses are very similar (both are $\sim 0.4\times10^9$ \msun) --- to the extent that the NW arm can be separated from the rest of the \hi\ disc in the $11''$-resolution cube.

\begin{figure}
\centering
\includegraphics[width=9cm]{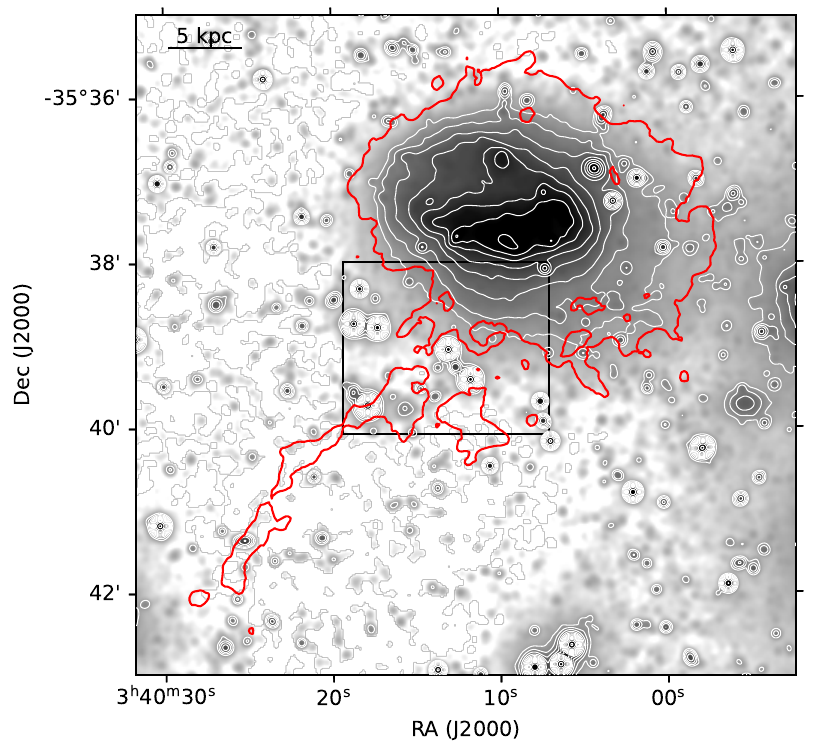}
\caption{Optical image ($g$-band, smoothed with a $5''$-FWHM Gaussian kernel) and its contours (in white) with, overlaid, the $11''$-resolution \hi\ contour at \Nhi\ $=1.5\times10^{20}$ cm$^{-2}$ (in red). The black box highlights the region where the southern \hi\ tail connects with the bump of the optical isophotes at the southern edge of the stellar body.}
\label{fig:bump}
\end{figure}

If the southern \hi\ tail and the NW \hi\ arm have a tidal origin and were later shaped by ram pressure, they might both have a stellar counterpart. \cite{leewaddell2018} discussed a possible counterpart of the southern tail based on the same optical images used in our work. They highlighted an optical bump at the southern edge of the stellar body, which might be a short stellar tidal tail indeed (see their Figure 2b). In their work, the positional match between this bump and the southern \hi\ tail was only approximate given the low angular resolution of their \hi\ data, but we can now confirm it with higher precision thanks to our $11''$-resolution \hi\ image (Fig. \ref{fig:bump}).

While Fig. \ref{fig:bump} strengthens the idea that the southern tail has a tidal origin and was later shaped by hydrodynamical forces that did not affect its stellar counterpart, we note that the optical images studied here show no obvious optical features corresponding to the NW arm. The northern part of the stellar body is far from settled as it hosts the northern plume and the northern clump, but we find no obvious connection to the NW \hi\ arm.

To conclude, adding to the above arguments the abundance of kinematically anomalous \hi\ within the main body of NGC~1427A (Figs. \ref{fig:t06chans}, \ref{fig:vf11} and \ref{fig:vdisp11}) and the peculiar velocity jumps of \hi\ along the major axis (Fig. \ref{fig:pvdcen}), we find that the hypothesis of a galaxy-galaxy interaction is strengthened by our new data. However, the details of the interaction remain an open question. In particular, both a high-speed interaction and a relatively advanced merger are plausible, but in neither case can we identify the interacting systems unambiguously. Additional work will be needed to reach a definitive conclusion.

\section{Summary}
\label{sec:summary}

We presented a study of the galaxy NGC~1427A observed as part of the MeerKAT Fornax Survey. Thanks to the excellent angular resolution (down to $\sim10''$ or $\sim1$ kpc), velocity resolution (1.4 \kms) and \Nhi\ sensitivity (down to $\sim10^{18}$ cm$^{-2}$) of our MeerKAT data we were able to study the distribution and kinematics of \hi\ in this galaxy in unprecedented detail. We found evidence that NGC~1427A is the remnant of a galaxy-galaxy interaction or merger and that it is subject to a blue-shifting and SE-bound ram-pressure wind caused by its motion through the ICM in the central region of the Fornax cluster. The key evidence in favour of ram pressure is: \it i) \rm the morphology and kinematics of the one-sided, starless, radially-oriented, southern \hi\ tail, which was initially discovered in previous work but we revealed here in its full (to date) size and complexity; and \it ii) \rm the large-scale distortion of the \hi\ velocity field in the outer region of the stellar body. The key evidence in favour of a recent galaxy-galaxy interaction or merger is: \it i) \rm the close match between the strongly disturbed distribution of stars and that of the dense \hi\ in the stellar body; \it ii) \rm the abundance of kinematically anomalous gas in the galaxy; \it iii) \rm the complex connection of the southern \hi\ tail with the main \hi\ body, ruling out a simple ram pressure origin of the tail; and \it iv) \rm the negligible impact of a tidal interaction with the cluster gravitational potential.
 
Based on our analysis, the strongest component of the ram-pressure wind felt by NGC~1427A because of its motion through the ICM is along the line of sight and directed towards us, consistent with the $\sim600$ \kms\ red-shift of the galaxy relative to the cluster centre. This component is sufficient to offset the velocity of the \hi\ outside the half-light radius ($\sim 5$ kpc) towards lower recessional velocities. The velocity offset grows larger with increasing radius, consistent with the decrease of the galaxy restoring force, and bends the kinematical minor axis into a U shape which we were able to model.

The same component of the ram-pressure wind also causes a velocity gradient along the extended, one-sided southern \hi\ tail. Our data increase the projected length and \hi\ mass of the tail by a factor of 3 and 50\%, respectively, bringing them to $\sim70$ kpc and $\sim0.4\times10^9$ \msun. Given the strong ram pressure along the line of sight, it is likely that the true length of the tail is even larger. Its considerable width of $\sim30$ kpc, which is comparable to the diameter of the main \hi\ body, is another strong indication that ram pressure is at work, as tidal tails are usually much narrower.

The tail owes its SE orientation to the component of the ram-pressure wind on the plane of the sky, which implies a radial infall of NGC~1427A towards the cluster centre located towards NW. Following the SE direction of the tail past its end and towards the outer regions of the cluster, we find a population of scattered, starless \hi\ clouds whose velocity is consistent with the velocity gradient along the tail. The clouds reach out to 300 kpc from NGC~1427A and could be the most ancient remnant of the passage of this galaxy through the ICM.

The SE-bound component of the ram-pressure wind is weaker than the one along the line of sight, too weak to cause a compression of the outer \hi\ disc in any direction on the plane of the sky. As in \cite{leewaddell2018}, we rule out the suggestion found in the literature that the SW star-forming front within the bright part of the stellar body is caused by ram pressure. That front is embedded deeply into the \hi\ body, with no indication of ISM compression, and must thus be caused by other mechanisms.

Our data bring new evidence that the galaxy, while being subject to the effects of ram pressure, has also been shaped by tidal forces. First, the highly disturbed distribution of dense \hi\ in the stellar body tracks closely the distribution of bright optical light, implying that the forces at play act on both gas and stars. Furthermore, while most of the \hi\ follows a relatively regular rotation pattern, we detect abundant kinematically anomalous \hi, too. Finally, the complex connection of the southern \hi\ tail with the inner \hi\ body, which includes the inversion of the tail's velocity gradient observed at larger radius, is inconsistent with a ram pressure origin. This suggests that the tail must have started during a tidal interaction, and was only later blue-shifted and stretched towards SE by ram pressure. Indeed, we strengthen previous results that the southern \hi\ tail has a stellar counterpart at its base, and that there might be a twin northern tail which was later blue-shifted and pushed back against the stellar body by ram pressure.

We rule out that the tidal interaction revealed by our data is with the overall cluster potential, as our estimate of the tidal radius is well outside the galaxy ($\sim20$ kpc). Therefore, we conclude that NGC~1427A is the remnant of a tidal interaction between two galaxies. We estimate the interaction to have started at least 300 Myr ago. The interaction could have occurred because of a high-speed encounter between NGC~1427A and another Fornax galaxy, for which we identify a few possible candidates. Alternatively, NGC~1427A is the result of a merger, where the two merging galaxies have already coalesced and are now in the process of settling within the remnant. The system is, however, too complex to reach a definitive conclusion on the details of the interaction.

Our results demonstrate that, in a small cluster like Fornax, galaxy interactions and mergers can enhance the efficiency of ram-pressure stripping. In the case of NGC~1427A we find that, because of its relatively low mass and high speed within the cluster, the ram pressure is able to displace the ISM located in the outer region of the stellar body. The interplay between tidal and hydrodynamical effects leads to a complex distribution and kinematics of the ISM, which is likely to have a long-term effect on the subsequent evolution of the galaxy.

\begin{acknowledgements}

The MeerKAT telescope is operated by the South African Radio Astronomy Observatory, which is a facility of the National Research Foundation, an agency of the Department of Science and Innovation. This project has received funding from the European Research Council (ERC) under the European Union's Horizon 2020 research and innovation programme (grant agreement no. 679627; project name FORNAX). PK is partially supported by the BMBF project 05A23PC1 for D-MeerKAT. DJP and NZ greatly acknowledge support from the South African Research Chairs Initiative of the Department of Science and Technology and National Research Foundation. Part of the data published here have been reduced using the CARACal pipeline, partially supported by ERC Starting grant number 679627, MAECI Grant Number ZA18GR02, DST-NRF Grant Number 113121 as part of the ISARP Joint Research Scheme, and BMBF project 05A17PC2 for D-MeerKAT. Information about CARACal can be obtained online under the URL: https://caracal.readthedocs.io. We acknowledge the use of the ilifu cloud computing facility, www.ilifu.ac.za, a partnership between the University of Cape Town, the University of the Western Cape, the University of Stellenbosch, Sol Plaatje University, the Cape Peninsula University of Technology and the South African Radio Astronomy Observatory. The Ilifu facility is supported by contributions from the Inter-University Institute for Data Intensive Astronomy (IDIA, a partnership between the University of Cape Town, the University of Pretoria, the University of the Western Cape and the South African Radio astronomy Observatory), the Computational Biology division at UCT and the Data Intensive Research Initiative of South Africa (DIRISA).

\end{acknowledgements}

%
%

\bibliographystyle{aa} 
\bibliography{myrefs} 

\end{document}